\documentclass[10pt,aps,pre,twocolumn,superscriptaddress,citesort]{revtex4-2}

\usepackage{graphicx}
\usepackage{amsmath}
\usepackage{amsfonts}

\usepackage{bm}
\usepackage{xcolor}
\newcommand{\SIapp}{\textcolor{blue}{\textit{SI Appendix}}}
\definecolor{tab_blue}{HTML}{1F77B4}
\usepackage{titlesec}

\usepackage[colorlinks=true,linkcolor=blue,citecolor=blue,urlcolor=blue]{hyperref}

\hypersetup{%
  breaklinks=true,
  colorlinks=true,
  linkcolor=tab_blue,
  filecolor=tab_blue,
  urlcolor=tab_blue,
  citecolor=tab_blue,
  pdfborder=0 0 0,
}

\begin{document}

\title{Avalanches of choice: how stranger-to-stranger interactions shape crowd dynamics}

\author{Ziqi Wang}
\affiliation{Fluids and Flows group and J.M. Burgers Center for Fluid Mechanics, Department of Applied Physics and Science Education, Eindhoven University of Technology, 5600 MB Eindhoven, Netherlands}
\author{Alessandro Gabbana}
\affiliation{Department of Physics and Earth Sciences, University of Ferrara, 44122 Ferrara, Italy}
\affiliation{INFN Ferrara, 44122 Ferrara, Italy}
\author{Federico Toschi}
\email{f.toschi@tue.nl}
\affiliation{Fluids and Flows group and J.M. Burgers Center for Fluid Mechanics, Department of Applied Physics and Science Education, Eindhoven University of Technology, 5600 MB Eindhoven, Netherlands}
\affiliation{Consiglio Nazionale delle Ricerche - Istituto per le Applicazioni del Calcolo, Rome I-00185, Italy}

\date{\today}

\begin{abstract}
Pedestrian routing choices play a crucial role in shaping collective crowd dynamics, yet the influence of interactions 
among unfamiliar individuals remains poorly understood. 
In this study, we analyze real-world pedestrian behavior at a route split within a busy train station using 
high-resolution trajectory data collected over a three-year time frame. 
We disclose a striking tendency for individuals to follow the same path as the person directly in front of them,
even in the absence of social ties and even when such a choice leads to a longer travel time.
This tendency leads to bursty dynamics, where sequences of pedestrians make identical decisions in succession,
leading to strong patterns in collective movement. 
We employ a stochastic model that includes route costs, randomness, and social imitation to accurately reproduce 
the observed behavior, highlighting that local imitation behavior is the dominant driver of collective routing 
choices. 
These findings highlight how brief, low-level interactions between strangers can scale up to influence large-scale 
pedestrian movement, with strong implications for crowd management, urban design, and the broader understanding 
of social behavior in public spaces.
\end{abstract}
\maketitle	

Pedestrian dynamics in crowded spaces emerge from individuals navigating toward defined goals and from
their adaptive behavior as they interact with one another and with their physical surroundings. 
These local interactions give rise to emergent collective behaviors such as lane formation, stop-and-go waves, and bottlenecks, 
which can significantly impact the efficiency and safety of movement through public 
infrastructure~\cite{corbetta2023physics, hoogendoorn2005pedestrian, kretz2006experimental,solmaz2012identifying}. 
From this point of view, routing choice is particularly consequential: although seemingly simple and straightforward, 
it emerges from a subtle and often competing interplay of travel time, path length, crowd density, directional changes, 
and perceived comfort~\cite{hughes2003flow,seneviratne-tpt-1985,hoogendoorn-tr-2004,brown-eb-2007,guo-jotg-2013,shatu-jotg-2019,basu2022systematic}.
Understanding how individuals make such decisions in real-world settings is therefore essential for the design and operation 
of transportation hubs, public buildings, and evacuation procedures. 

Quantitatively, routing choice is often formalized through cost-minimization models, where individuals are assumed to select paths that minimize a perceived travel cost function. 
These models successfully reproduce aggregate routing patterns under few specific configurations~\cite{borgers1986model,van2014using,hoogendoorn2004pedestrian}, 
but typically treat pedestrians as independent decision-makers, unless they belong to predefined social groups~\cite{moussaid2010walking,cheng2014review}. 
However, growing empirical and theoretical evidence shows that local interactions, particularly imitation and social influence, significantly affect the collective crowd dynamics. For instance, herding tendencies at bottlenecks and junctions have been observed in both laboratory and real-life scenarios~\cite{kretz-josm-2006,moussaid-pnas-2011,gabbana2022fluctuations,gabbana2021modeling}, 
and can be modeled effectively by introducing imitation rules into routing frameworks~\cite{gabbana2022fluctuations,gabbana2021modeling}.

Within this scenario, recent research has highlighted the follow-the-leader (L-F) effect, 
where individuals (followers) tend to align with those who first initiate movement (leaders)
\cite{dyer2009leadership,pelechano2006modeling,boos2014leadership}, as a key behavioral mechanism.
Experiments have also shown that individuals are more likely to follow nearby individuals than distant ``opinion leaders'' \cite{ding2020experimental},
and that the characteristics of leaders, such as their speed, direction, and location relative to followers, 
can significantly impact crowd movement and evacuation efficiency in emergency drills~\cite{zhang2021experimental,dyer2009leadership,qingge2007simulating}.
Recent work has further explored dynamic L-F interactions, wherein leader-follower roles emerge spontaneously and adaptively, 
allowing individuals to switch whom they follow based on evolving spatial-temporal context~\cite{xie2022simulation,yu2025study}.

While these studies underscore the importance of local imitation, they also face several critical limitations:
(i) participants in controlled experiments are often aware of the study purpose, introducing subjective bias;
(ii) laboratory settings are highly confined and simplified, lacking the spatial complexity of real facilities; (iii) sample sizes are typically on the order of a few dozen to a hundred individuals, limiting statistical robustness; and (iv) most experiments focus on emergency scenarios, offering limited insight into everyday navigation.

Most importantly, previous studies have rarely distinguished between following behavior among socially connected individuals (e.g. friends, families) and those among complete strangers. This distinction is crucial: it remains unclear whether, and to what extent, strangers influence each other's routing decisions in non-emergency, real-life conditions.
Addressing this question is challenging due to the difficulty of observing spontaneous interactions in realistic settings with natural flow, spatial complexity, and no experimental manipulation. 

Progress is further constrained by the technical challenge of obtaining long-term, high-resolution measurements in real-life facilities. Consequently, most research on crowd dynamics \cite{haghani2018crowd} relies heavily on qualitative simulations \cite{cristiani2019robust,dang2024literature} based on microscopic \cite{helbing1995social,blue1998emergent,blue2001cellular,helbing2000simulating,kirchner2002simulation} or macroscopic models \cite{hughes2000flow,hughes2002continuum,treuille2006continuum,duives2013state}, or on controlled experiments \cite{kretz2006experimental,moussaid2011simple,cao2017fundamental,moussaid2011simple} and pedestrian interviews \cite{borgers1986model,verlander1997pedestrian,koh2013influence}, all of which inevitably alter the natural dynamics under observation. 
Consequently, the effect of stranger interactions on pedestrian routing decisions, which is an essential component for understanding the variability and unpredictability of real-world crowd flows, remains largely unexplored.

In this study, we bridge these gaps by leveraging a large-scale, high-resolution dataset collected at Eindhoven Centraal Railway Station (The Netherlands), thanks to the advances in overhead depth sensing which provides precise, continuous monitoring of pedestrian movements in (even high density) real-world settings without compromising privacy \cite{brvsvcic2013person,corbetta2014high,seer2014kinects,kroneman2018accurate,willems2020pedestrian}. Over a period from March of 2021 to March of 2024, individual trajectories occurring on the station platform were recorded with millimeter accuracy using a commercial overhead tracking system. From this dataset, approximately 100,000 passengers who exited the train through specific door locations were selected for analysis (see Methods for experimental campaign details). 
This unprecedented, high-statistics dataset enables the reconstruction of pedestrian movements with high temporal ($\approx$ 0.1~s) and spatial ($\approx$ 0.001~m) resolution, offering a real-world statistical foundation that goes far beyond laboratory-scale experiments.


In this framework, we focus on a subset of pedestrian trajectories, where individuals, after de-boarding a train, face a binary routing choice between a direct, shorter path and a longer path that involves circumventing a kiosk in the middle of the platform.
This simple scenario serves as a foundational model for studying pedestrian routing behavior.
Through our high-resolution trajectory measurements, statistical analysis, and stochastic routing modeling, 
we uncover a strong ``stranger-following effect'': even when traveling alone (i.e., not part of a social group), 
pedestrians show a strong tendency to select the same path as the person immediately ahead of them. 

These dynamics exhibit burstiness and temporal clustering: once a path becomes locally favored, subsequent
individuals disproportionately follow it, leading to temporary flow imbalances. Notably, this behavior does not
optimize throughput or individual efficiency. While system-wide throughput would be maximized if pedestrians were to split
evenly between the two paths, we find that imitation-driven choices often create suboptimal routing distributions, both for
individuals and for the collective.

By incorporating an imitation mechanism into a routing model, we show that stranger-following dominates over classical 
factors such as herding and walking-speed variability, providing a new explanation for how microscopic fluctuations scale into macroscopic flow patterns. 
Unlike classical leader-following behavior, where individuals intentionally follow a designated or knowledgeable leader (e.g., trained personnel or visibly informed agents), the imitation observed here occurs among anonymous pedestrians with no social hierarchy or communication. In our setting, pedestrians lack prior knowledge of which route is preferable, and thus face decision uncertainty. This inference is transient and purely local, yet it leads to stranger-to-stranger imitation that differs fundamentally from leader-driven following and can still create collective route-choice cascades.

Our study revises the standard picture of pedestrian routing choice by identifying stranger influences as a robust and quantifiable driver of crowd collective dynamics, highlighting how spontaneous imitation can prevent the system from achieving either the individual or the facility-optimal travel times, with implications for real-time crowd management, emergency planning, and urban design.

\section*{Overview of path choices}
We first examine the overall patterns of pedestrian routing choice. 

\begin{figure}[t!]
\centering
\includegraphics[width=1\linewidth]{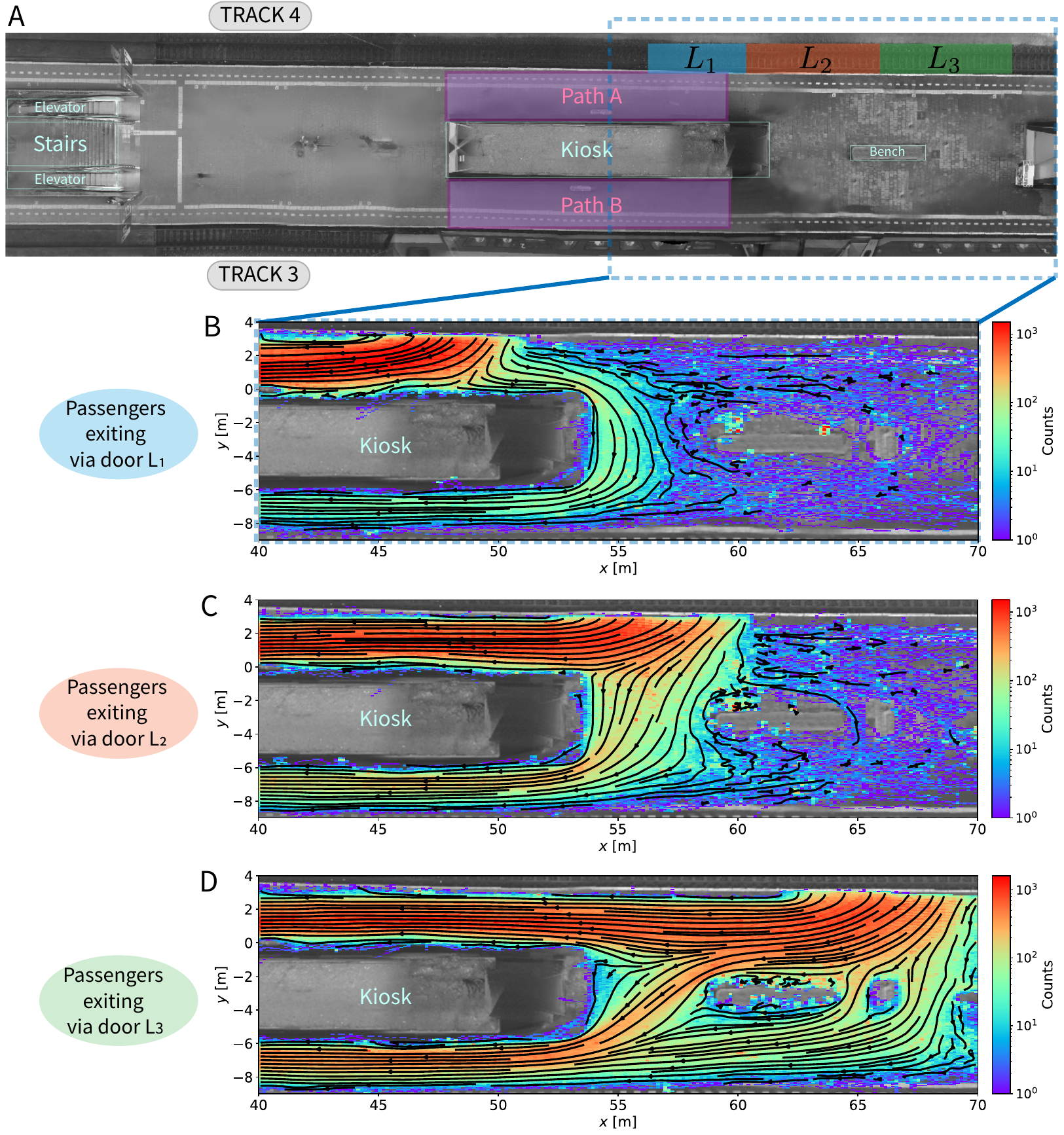}
\caption{\textbf{Real-life measurement setup and pedestrian flow patterns at Eindhoven Centraal Railway Station (The Netherlands)}.
(A) Experimental layout of the platform. Passenger trajectories are recorded using a commercial tracking system. Disembarking passengers from Track 4 exit the train via three door locations---$L_1$, $L_2$, and $L_3$---and encounter a path bifurcation, choosing between Path A and Path B. A large indoor waiting area (Kiosk) is present on the platform, along with other marked elements such as elevators, staircases, and benches.
Heat maps of pedestrian positions after exiting through doors (B) $L_1$, (C) $L_2$, and (D) $L_3$, respectively. Colors represent the density of observed positions on a logarithmic scale based on a 2D count histogram. Overlaid streamlines show the mean velocity vector field (spatially binned), highlighting the most probable pedestrian trajectories.
}
\label{FIG1}
\end{figure}

Due to variability in train stopping positions, the relative location of each door to the exit varies from train to train.
To capture the range of possible passenger disembarking positions, we focus on a scenario (Fig.~\ref{FIG1}A) in which
passengers disembark from three representative door zones, $L_1$, $L_2$, and $L_3$ (see Methods).
These locations span increasing distances from a bifurcation point where the platform path splits into two 
options for reaching the station exit:
(i) Path A: a direct, shorter route, and
(ii) Path B: a slightly longer route that loops around a central kiosk.
For each pedestrian $i$, we record their routing choice $X_i \in \{A,B\}$,
and their sequence index $i$, i.e. the relative order in which they exited the train.
This setup provides a natural testbed for studying how individuals and crowds resolve routing choices 
in the presence of congestion, spatial heterogeneity, and local social dynamics, especially among strangers.

Figure~\ref{FIG1}(B-D) shows the resulting pedestrian flow fields. Density heatmaps reveal strong spatial heterogeneity across different door positions: trajectories from $L_1$ and $L_2$ are more strongly concentrated near Path A (Fig.~\ref{FIG1}B and C), whereas flows from $L_3$ exhibit broader dispersion and a more balanced usage of both paths (Fig.~\ref{FIG1}D). 
While spatial constraints and distance clearly play a role, these factors alone cannot explain why pedestrians arriving under similar conditions sometimes make different path choices. This raises a key question: to what extent are individual routing decisions shaped by the behavior of others, in particular by strangers?

\begin{figure*}[t!]
\centering
\includegraphics[width=0.99\linewidth]{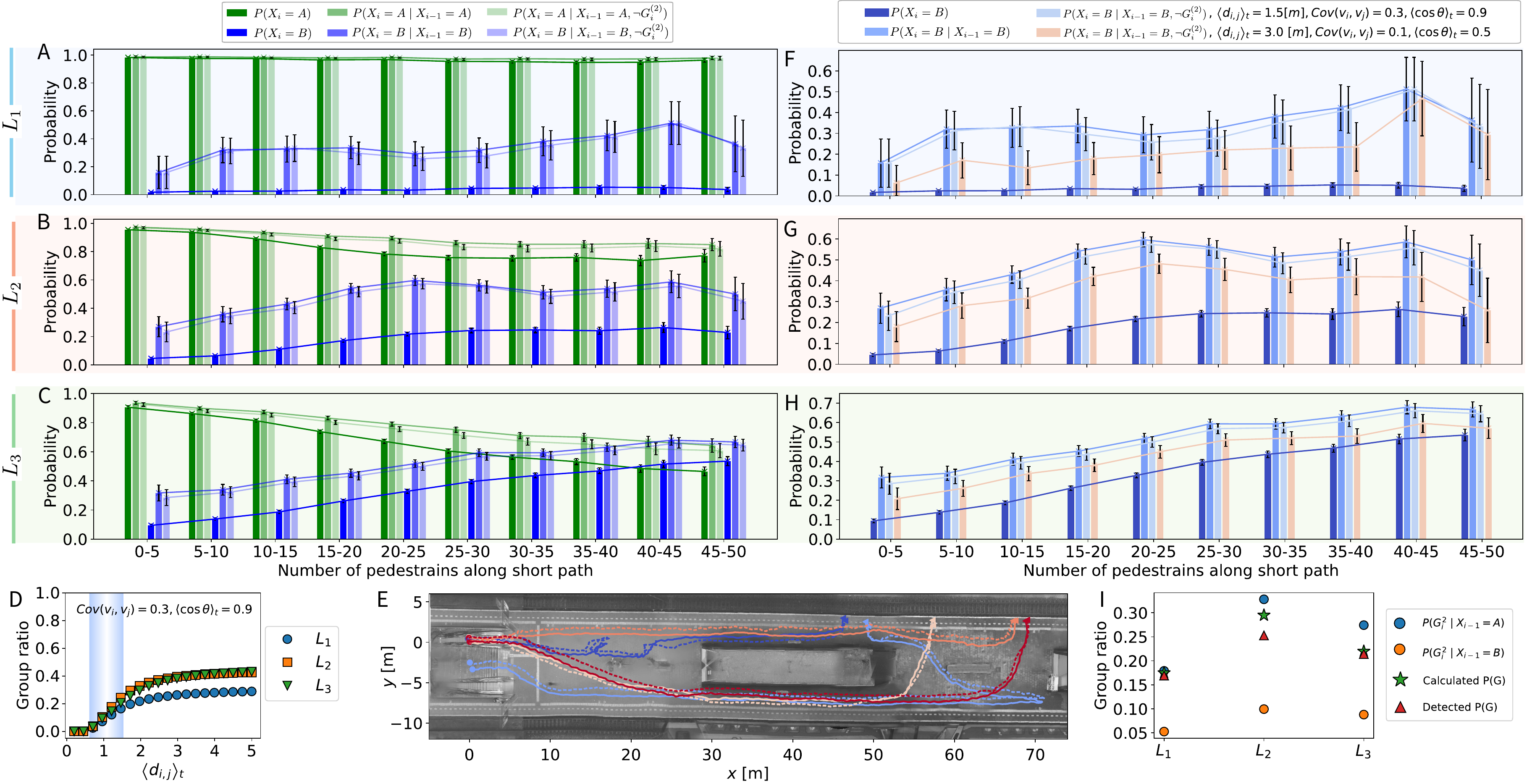}
\caption{\textbf{Statistical analysis of pedestrian path-choice behavior.}
(A-C) Path-choice probabilities for passengers exiting through doors (A) $L_1$, (B) $L_2$, and (C) $L_3$, plotted against the number of pedestrians on the shorter path (Path A). We report the marginal probability $P(X_i=\chi)$, the conditional probability $P(X_i=\chi|X_{i-1}=\chi)$, and the probability excluding groups $P(X_i=\chi|X_{i-1}=\chi, \neg G_i^{(2)})$, with $\chi=A$ (green colors) and $\chi=B$ (blue color).
The observation of $P(X_i=B|X_{i-1}=B,\neg G_i^{(2)})) \approx P(X_i=B|X_{i-1}=B) > P(X_i=B)$ shows a clear evidence of stranger-following effect.
(D) Group ratio $P(G)$ as a function of the distance threshold with covariance of speeds $\mathrm{Cov}(v_i,v_j)\!>\!0.3$ and mean directional alignment $\langle \cos\theta\rangle_t\!>\!0.9$, for door position of $L_1$ (blue), $L_2$ (orange), and $L_3$ (green), respectively.
In this work, the group detection are based on the thresholds $\left\langle d_{i,j} \right\rangle_t = 1.5~[m]$, $\mathrm{Cov}(v_i,v_j)=0.3$, and $\left < \cos \theta \right >_t = 0.9$. The corresponding example trajectories of pedestrians identified as belonging to the same group (same color; solid vs.\ dashed for different members) are shown in panel (E), with triangles and circles marking start and end points, respectively. For simplicity, we present only groups of size two, which make up the majority of all observed groups.
(F-H) Robustness of the stranger-following effect: $P(X_i{=}B \mid X_{i-1}{=}B,\, \neg G_i^{(2)})$ computed under two sets of thresholds for the group criteria: (light blue) a reasonable set ($\langle d_{i,j}\rangle_t{=}1.5$\,m, $\mathrm{Cov}(v_i,v_j){=}0.3$, $\langle \cos\theta\rangle_t{=}0.9$) and (red) a deliberately lenient set ($\langle d_{i,j}\rangle_t{=}3.0$\,m, $\mathrm{Cov}(v_i,v_j){=}0.1$, $\langle \cos\theta\rangle_t{=}0.5$). The relation $P(X_i{=}B \mid X_{i-1}{=}B,\, \neg G_i^{(2)}) \approx P(X_i{=}B \mid X_{i-1}{=}B) > P(X_i{=}B)$ persists, indicating robustness to reasonable variations in detection thresholds, provided the threshold is within a reasonable range and applied uniformly across door locations.
(I) Comparison of group ratios obtained in three ways for each door: (i) conditioned estimates from path-choice statistics $P(G_i^{(2)}\mid X_{i-1}{=}A)$ (blue) and $P(G_i^{(2)}\mid X_{i-1}{=}B)$ (orange), (ii) the resulting weighted estimate $P(G)$ (green), and (iii) direct detection based on the three criteria (red). Overall path-choice probabilities $P(X_i{=}A)= 0.967, 0.855, 0.707$ for $L_1$, $L_2$, and $L_3$, respectively.
}
\label{FIG2}
\end{figure*}
\section*{The follow-the-leader effect: beyond social groups}

A central focus of our study is whether pedestrians are influenced by the path
choices of others, beyond the well-documented effect of traveling in groups \cite{moussaid2010walking, pouw2020monitoring}. 

To disentangle these effects, we classify each $i^{th}$ passenger disembarking from the train as either: an individual
($I_i$), a group leader ($G_i^{(1)}$), or group members ($G_i^{(2)}$). We assume that group members (one or many) follow the path choice of their leader. 
We define the order of the passenger based on ordered exit times measured within a two-meter detection zone around each door location (\SIapp).

In order to quantify path-choice behavior, we introduce the definition of three types of probabilities:
(i) marginal probability $P(X_i = \chi)$, giving the overall baseline likelihood that a pedestrian chooses path $\chi \in \{A, B\}$;
(ii) conditional probability $P(X_i = \chi \mid X_{i-1} = \chi)$, measuring the tendency of a passenger to repeat the choice made by the immediately preceding one, thereby quantifying potential short-term social influence;
(iii) conditional probability excluding groups $P(X_i = \chi \mid X_{i-1} = \chi, \neg G_i^{(2)})$, which is restricted to the sole pedestrians not identified as group members, thus isolating stranger-to-stranger effects.

\subsection*{Group detection methodology}
To reliably distinguish true groups from independent pedestrians, we developed a group detection method based on three criteria: average coexistence distance $\left\langle d_{i,j} \right\rangle_t$ (Methods and \SIapp), speed covariance $\mathrm{Cov}(v_i,v_j)$, and directional alignment $\langle \cos\theta\rangle_t$ ($\theta$ is the angle between $\vec{v}_i$ and $\vec{v}_j$, where subscript $i$ and $j$ denote the $i^{\text{th}}$ and $j^{\text{th}}$ pedestrians, respectively), and $\langle \cdot \rangle_t$ is the time average.

The detected group ratio $P(G)$ is generally robust to reasonable variations of the detection thresholds. When all three criteria are combined, $P(G)$ remains stable across different door locations within a broad range of distance thresholds (0.5-2.0\,m) which is reasonable because group composition is intrinsic and largely independent of exit door location, as illustrated in Fig.~\ref{FIG2}D for $\mathrm{Cov}(v_i,v_j)\!>\!0.3$ and $\langle \cos\theta\rangle_t\!>\!0.9$.
The detected groups composed of one member and one leader are the most frequent \cite{moussaid2010walking}. The group size distribution, detailed detection methods, and analyses of $P(G)$ for each criterion are provided in \SIapp.

\subsection*{Experimental evidence} 
With these definitions in place, we now examine the empirical results for passengers exiting from doors $L_1$
(Fig.~\ref{FIG2}A), $L_2$ (Fig.~\ref{FIG2}B), and $L_3$ (Fig.~\ref{FIG2}C).
The results show a clear sequential dependence: the probability of selecting a path is systematically larger when the preceding pedestrian made the same choice, i.e. $P(X_i=\chi \mid X_{i-1}=\chi) > P(X_i=\chi)$. 
Importantly, this effect persists even when groups are excluded, with $P(X_i=\chi \mid X_{i-1}=\chi, \neg G_i^{(2)}) \approx P(X_i=\chi\!\mid\!X_{i-1}=\chi)\!>\!P(X_i=\chi)$ and remains robust under all observed congestion levels on the shorter path (i.e.,  Path A).
These observations demonstrate that routing choices are not independent: passengers tend to follow the decision of the person in front, even when no social ties are present. We refer to this as the stranger-following effect, revealing that even strangers exert measurable influence on each other's path selection.

For the conditional probabilities shown in Fig.~\ref{FIG2}(A-C), group detection used thresholds $\left\langle d_{i,j} \right\rangle_t = 1.5~[m]$, $\mathrm{Cov}(v_i,v_j)=0.3$, and $\left < \cos \theta \right >_t = 0.9$, based on which example trajectories of pedestrians identified as belonging to the same group are illustrated in Fig.~\ref{FIG2}E (same color, solid vs. dashed lines, indicates the same group).

A natural question arises: does the stranger-following effect change with different group detection thresholds? To test this, we performed robustness checks by using two sets of thresholds: a reasonable set (light blue) and a deliberately lenient set (red), as shown in Fig.~\ref{FIG2}(F-H). Across different door locations, the relation $P(X_i{=}B \mid X_{i-1}{=}B,\, \neg G_i^{(2)}) \approx P(X_i{=}B \mid X_{i-1}{=}B) > P(X_i{=}B)$ persists, which confirms that the stranger-following effect is not an artifact of group misclassification. Even under deliberately lenient thresholds, the conditional probability remains elevated, which again demonstrates that pedestrians tend to imitate the path of the person directly in front of them, reflecting a localized form of social influence on decision-making.

\subsection*{Inferring group ratio from path-choice statistics}
An important implication of the statistical characteristics shown in Fig.~\ref{FIG2}(A-C) is that the group ratio can be inferred directly from path-choice behavior. 
It can be shown (\SIapp) that the probability of a pedestrian being group members, $P(G_i^{(2)})$, can be estimated directly from the experimentally measured conditional path-choice probabilities.

Specifically, the group ratio can be inferred as 
\begin{equation}
\label{eq:overall_group_ratio}
P(G) = \sum_{\chi \in \{A, B\}} 2P(G_i^{(2)} \mid X_{i-1}=\chi) \cdot P(X_i=\chi).
\end{equation}
where $P(X_{i}=\chi)$ are obtained from experimental results. The conditioned probability, $P(G_i^{(2)} \mid X_{i-1}=\chi)$ (circles in Fig.~\ref{FIG2}I), can be directly measured from the experimental data (Fig.~\ref{FIG2}(A-C)).

As shown in Fig.~\ref{FIG2}J, the $P(G)$ inferred from path-choice statistics (green stars) agrees remarkably well with that detected from trajectory-based group identification (red triangles), with a discrepancy below 0.05. This confirms that the group ratio can be reliably estimated from path-choice statistics.
This establishes a quantitative link between microscopic path-choice correlations and macroscopic group composition.

\begin{figure*}[t!]
\centering
\includegraphics[width=.99\linewidth]{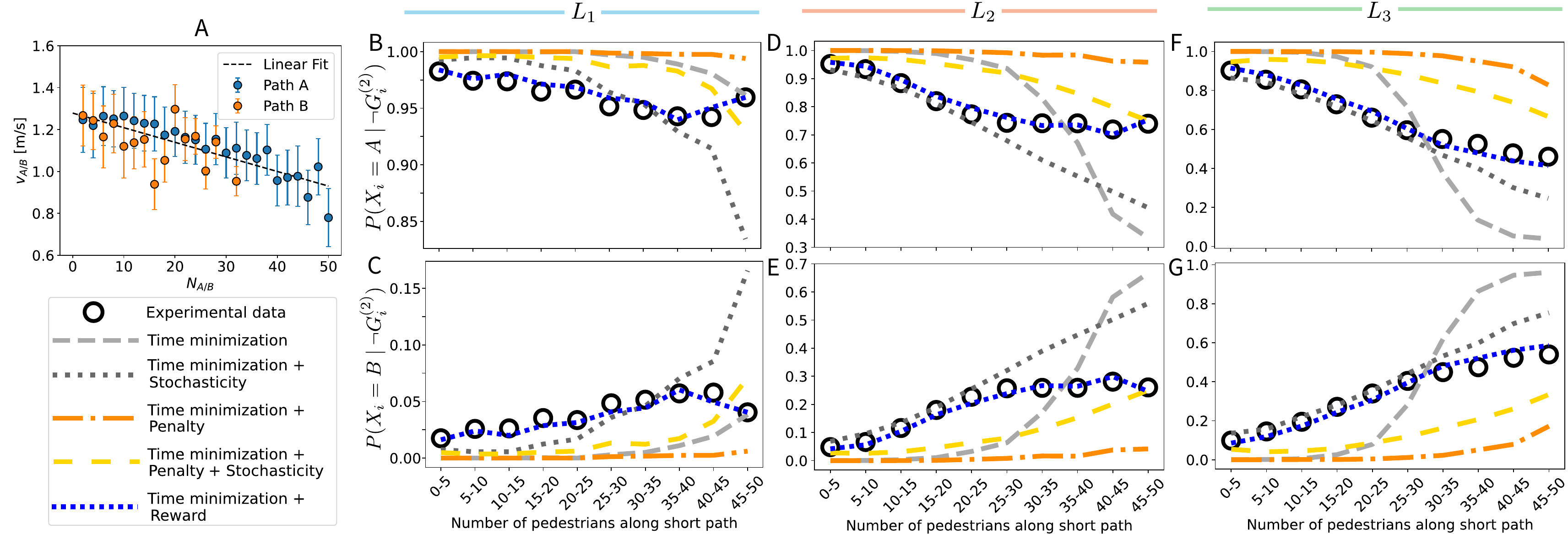}
\caption{\textbf{Theoretical routing model results of path choice.} 
(A) Fundamental diagram showing the relationship between the average walking speed and the number of pedestrians observed within the same fixed-size segment (purple-shaded regions in Fig.~\ref{FIG1}A) on Path A ($N_A$, orange circles) and Path B ($N_B$, blue circles). The two regions have identical length and width; therefore, the counts directly represent comparable local crowding conditions. The dashed black line represents a linear fit of the form $V_i = V_0 - \kappa N_i, \quad i \in \{A, B\},$, with $V_0 = -0.007$ m/s and $\kappa = 1.2785$ m/s. Error bars denote standard deviations.
(B-G) Model-data comparison of the conditional probability excluding groups of choosing Path A (panels B, D, F) and Path B (panels C, E, G) as a function of the number of pedestrians on Path A, for exit doors $L_1$, $L_2$, and $L_3$.
Black symbols: experiments. 
Grey dashed: time-minimization-only baseline ($t_i$). 
Grey dotted: time-minimization-only $+$ speed variability ($S_i \cdot t_i$). 
Orange dash-dotted: herding only ($f_i \cdot t_i$).
Yellow dashed: herding $+$ speed variability ($r_i^{(\chi)} \cdot S_i \cdot t_i$).
Blue dotted: following stranger only ($r_i^{(\chi)} \cdot t_i$).
The following-stranger-only model (blue) already captures the main experimental trends, while adding herding and speed variability provide only minor local adjustments, showing that the stranger-following effect is an essential component of pedestrian decision-making. Only the following-stranger-only model (blue) is shown here for clarity; additional models that include stochasticity and/or penalty terms are provided in \SIapp. 
For all the modeling results shown in (B-G), (i) the penalty function in Eq.~\ref{eq:penalty} is with a same set of parameters (we apply the penalty function only to $T_i$ on Path A), i.e., $a=0.1, s=20, n_0=0.2, f_A=1$; (ii) the reward function in Eq.~\ref{eq:reward} is with the parameter $\Gamma = 0.9$.  
}
\label{FIG3}
\end{figure*}

\begin{figure*}[!t]
\centering
\includegraphics[width=.89\linewidth]{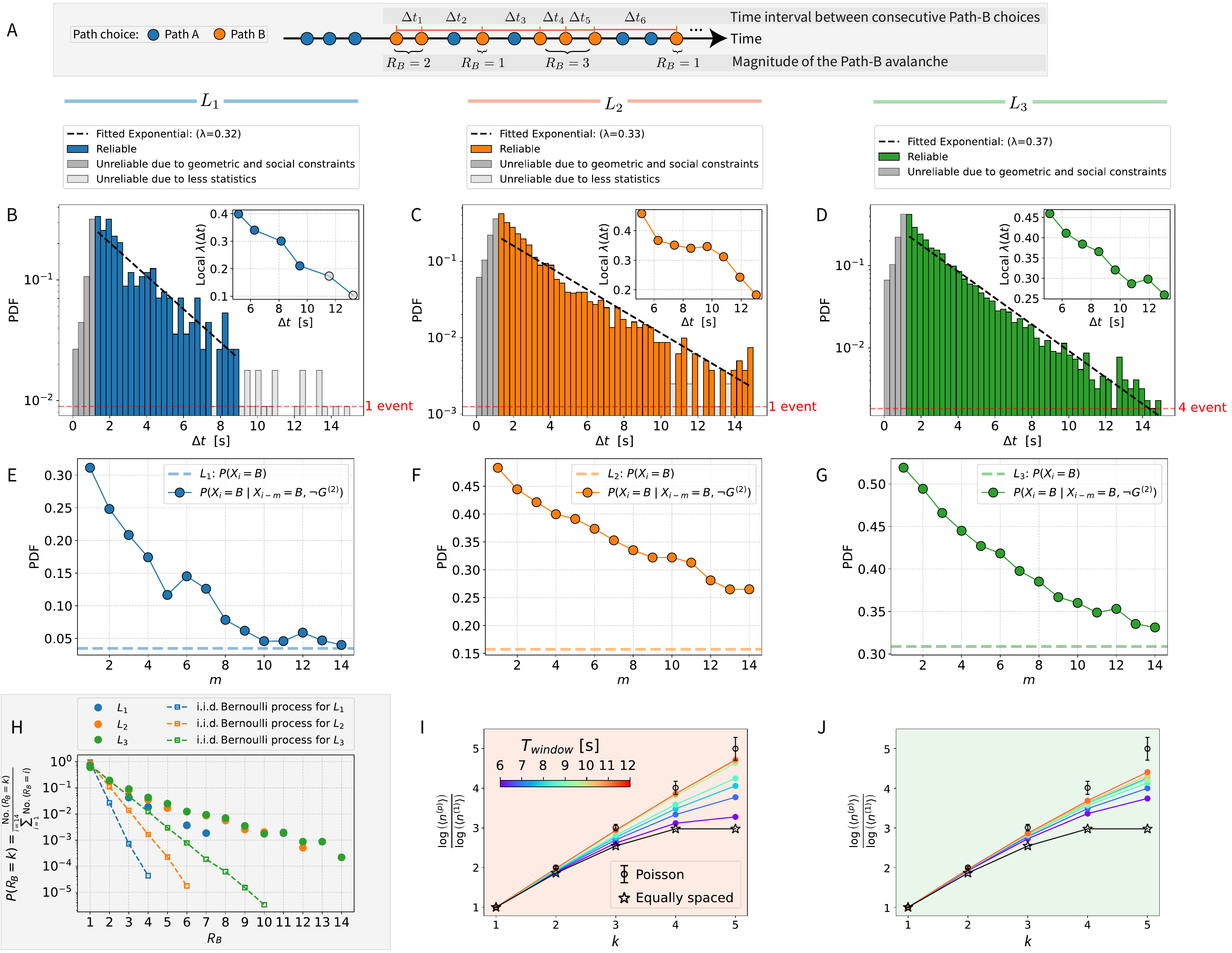}
\caption{\textbf{Avalanche of path choice.} 
(A) Definition of (i) the inter-person time interval, $\Delta t$, between consecutive pedestrians who choose Path B; and (ii) the magnitude of Path-B avalanche, $R_B$, defined as the number of consecutive pedestrians who choose Path B.
(B-D) Probability density function (PDF) of $\Delta t$ for doors $L_1$, $L_2$, and $L_3$, respectively. 
Black dashed lines show the exponential fit $f(\Delta t) = C e^{-\lambda \Delta t}$, where $\lambda$ is the average rate of choosing Path B. The fit excludes unreliable bins: (i) very small time intervals (typically $\Delta t<1.2$s), where the exponential trend cannot extend due to geometric or social constraints (gray bars), and (ii) insufficient statistics (light-gray bars).
Insets: corresponding effective local decay rate $\lambda(\Delta t)$. The rate is obtained from sliding-window exponential fits of the empirical PDF. Each point corresponds to a local regression over a 7.5 s window (20 bins), shifted by 1.13 s (detailed calculation process is reported in \SIapp). The deviation of $\lambda(\Delta t)$ from a constant value indicates that the underlying choice dynamics do not follow a uniform Poisson process.
The total number of events for analysis is 903 ($L_1$), 4182 ($L_2$), and 10529 ($L_3$).
(E-G) Conditional probability excluding group pedestrians, $P(X_i=B \mid X_{i-m}=B, \neg G^{(2)})$, for door location of $L_1$ (E), $L_2$ (F), and $L_3$ (G). The dashed lines indicate the probability of choosing Path B, $P(X_i=B)$. 
The fact that $P(X_i=B \mid X_{i-m}=B, \neg G^{(2)})$ remains systematically higher than $P(X_i=B)$ indicates that the Path-B choice influences persist across multiple individuals afterwards.
(H) The probability of an avalanche magnitude $R_B=k$ for door locations of $L_1$ (blue), $L_2$ (orange), and $L_3$ (green). Circles indicate the empirical results, while squares denote the corresponding i.i.d. Bernoulli baseline.
The probability $P(R_B=k)$ is computed as
$P(R_B=k) = \frac{\text{No}.(R_B=k)}{\sum_{i=1}^{i=14} \text{No}.(R_B = i)},$ where $\text{No}.(R_B=i)$ denotes the number of occurrences with $R_B=i$. The corresponding i.i.d Bernoulli process is generated by assigning each individual to choose Path B with probability $P(X_i=B)$ and Path A with probability $(1-P(X_i=B))$. For door location ($L_1-L_3$), the value of $P(X_i=B)$ is matched to its empirical counterpart (dashed lines in (E-G)). 
The empirical distributions of $R_B$ exhibit consistently heavier tails than those of an i.i.d. Bernoulli process, confirming the temporal clustering of choices for Path B: once one pedestrian chooses Path B, subsequent pedestrians are more likely to follow, forming ``avalanches'' of consecutive choices.
(I-J) Normalized factorial-moments analysis for $L_2$ and $L_3$ using sliding time windows of length $T_{\text{window}}$ (colorbar). We plot $\frac{\log\langle n^{(p)} \rangle}{\log\langle n^{(1)} \rangle}$ versus the order $p$, where $n^{(p)}$ is the $p$-th factorial moment of the event counts within each window. Results are compared with two benchmarks: a Poisson process with the same mean rate (black circles; average over 30 realizations, errorbars indicate standard deviation) and an artificially constructed equally spaced sequence with mean spacing comparable to the data (stars). 
No factorial moments are shown for $L_1$ due to the small number of Path B choices at this door, which prevents reliable estimation of higher-order statistics.
}
\label{FIG4}
\end{figure*}

\section*{Theoretical routing model}
To explain pedestrians (non-group members) path-choice behavior and the observed stranger-following effect, we develop a cost-based routing model \cite{gabbana2021modeling} in which each pedestrian selects the path with minimal perceived cost $T_i$.
The model builds on the empirical relation between walking speed and crowding, and incorporates two key behavioral mechanisms: (i) Social influence, which includes both herding (higher perceived cost for a less frequently chosen path) and the stranger-following effect (a local bias to repeat the previous pedestrian's choice); (ii) Stochasticity, to capture random variability in individual speed.

For each path $i \in \{A,B\}$, the average walking speed $V_i$ declines approximately linearly with the number of pedestrians $N_i$ counted within equally sized observation zones (purple-shaded regions in Fig.~\ref{FIG1}A) on the path $i$, as shown by the fundamental diagram in Fig.~\ref{FIG3}A. Since these zones have identical areas, the counts $N_i$ are proportional to local density, allowing for valid comparisons of congestion effects between different paths (\SIapp).
There are no significant differences in the walking speed across Path A and Path B, as shown by the probability density functions (PDFs) of $V_i$ 
reported in \SIapp. For simplicity, we consider a linear form $V_i = V_0 - \kappa N_i$ that captures this relationship for both paths, where $V_0$ denotes the free-walking speed and $\kappa$ is a congestion coefficient. This relation (Fig.~\ref{FIG3}A, dashed line) provides a predictive baseline for path-dependent travel time $t_i=l_i/V_i$, with $l_i$ as the mean streamline length connecting the corresponding door region to the selected path, computed from the spatially binned mean velocity field (details in \SIapp). 

At the bifurcation, each passenger selects a path by minimizing a perceived cost
\begin{equation}
T_i = f_i \cdot r_i^{(\chi)} \cdot S_i \cdot t_i,
\end{equation}
where three behavioral modifiers modulate the pure time minimization: (i) $S_i$ introduces speed variability through a stochastic realization of $V_i$; (ii) $f_i$ is a herding penalty that penalizes minority choices (and thus higher perceived cost) according to the instantaneous fraction of pedestrians on each path; (iii) $r_i^{(\chi)}$ rewards alignment with the pedestrian immediately ahead (and thus a reduced perceived cost), effectively biasing decision-making toward local consistency and transient stranger-following dynamics. Although more complex formulations are possible, the simplicity of our current model allows for a clear and interpretable decomposition of the contribution from the local interaction factor. 

Functional forms, parameter estimation, and model calibration of the behavioral modifiers are provided in the Methods.
This cost model links microscopic imitation and herding to macroscopic flow patterns and enables quantitative predictions of path-use ratios under different crowding conditions.

We tested the contributions of these behavioral modifiers by systematically comparing the conditional probability excluding the group, $P(X_i=\chi \mid \neg G_i^{(2)}), \ \chi \in \{A, B\}$ , from model predictions against experimental data (Fig.~\ref{FIG3}(B-G)). The baseline time-minimization model ($t_i$ only, gray dashed line) captures the broad effect of congestion but fails to reproduce the empirical transitions in real choice probabilities (symbols). Adding stochastic speed variability ($S_i t_i$, gray dotted line) improves the local fit by broadening the response but does not capture the full trend. Similarly, the penalty-only model from the herding effect ($f_i t_i$, orange dash-dotted line) remains insufficient to explain the observed path selection patterns. In contrast, the reward term from the stranger-following effect ($r_i^{(\chi)} t_i$, blue dotted line) alone recovers the primary structure of the empirical choice distributions, strongly suggesting that local imitation of immediate predecessors is the dominant driver of routing decisions under such configurations. Although speed stochasticity and penalty costs have been identified as influential factors in other contexts \cite{gabbana2021modeling,gabbana2022fluctuations}, their inclusion here yields only minor effects and leaves the qualitative behavior unchanged, indicating that their impact is subdominant in our setting (\SIapp). 

Taken together, these results reveal that pedestrian route choice at bifurcations is not fully captured by congestion-driven time minimization alone. Instead, social influences operating at different scales play decisive roles: immediate imitation of the preceding individual provides the strongest bias, while majority-following (herding) and speed stochasticity supply additional minor corrections. This hierarchical structure of influences highlights the importance of local, short-range interactions in shaping collective pedestrian flow patterns, with broader overall flowing tendencies emerging as secondary effects.  

\section*{Avalanche of path choice} The presence of stranger-following behavior suggests that routing decisions may be
temporally clustered, rather than occurring independently. To investigate this, we analyzed the distribution of inter-person intervals, $\Delta t$, which represent the time between consecutive pedestrians selecting Path B (Fig.~\ref{FIG4}A). 

If pedestrian decisions were independent, the inter-arrival times $\Delta t$ between consecutive selections of Path B would follow an exponential distribution, $P(\Delta t) = C e^{-\lambda \Delta t}$, with a constant decay rate $\lambda$, as expected for a Poisson process. 
However, in Fig.~\ref{FIG4}(B-D) insets, the fact that the local decay rate $\lambda(\Delta t)$ varies with $\Delta t$ shows that the underlying choice dynamics deviate from a uniform Poisson process.

To provide clearer evidence for the existence of choice avalanches, we computed the conditional probability $P(X_i=B \mid X_{i-m}=B, \neg G^{(2)})$. Physically, this quantity measures the tendency of pedestrian $i$ to follow the Path-B choice made by a stranger $(i-m)$ who is located $m$ decisions earlier, thereby capturing the persistence of short-range imitation along the decision sequence.
As shown in Fig.~\ref{FIG4}(E-G), for all door locations ($L_1$–$L_3$), $P(X_i=B \mid X_{i-m}=B, \neg G^{(2)})$ remains systematically higher than $P(X_i=B)$, which demonstrates that the influence of a prior Path-B choice propagates across multiple individuals downstream. The extent to which this influence persists, namely the magnitude of the choice avalanche, can be quantified using $R_B$, defined as the number of consecutive pedestrians choosing Path B (run-length of Path B).
Figure~\ref{FIG4}(H) shows the probability $P(R_B=k)$ for the empirical data (circles) together with a baseline generated from an independent and identically distributed (i.i.d.) Bernoulli process (squares), in which each pedestrian chooses Path B with the same empirical probability. The empirical distributions exhibit consistently heavier tails than those of the i.i.d. baseline, which confirms the temporal clustering of choices for Path B: once one pedestrian chooses Path B, subsequent pedestrians are more likely to follow, forming ``avalanches'' of consecutive choices. 

To further quantify these temporal correlations, we performed a factorial-moment analysis. For a Poisson process with
mean rate $\lambda$, the factorial moment of order $k$ for the event counts $n$ satisfies 
$ \left< n^{(k)} \right> = \lambda^k$ (details in \SIapp). As a consequence, the normalized ratio 
$\log\langle n^{(k)}\rangle / \log\langle n^{(1)}\rangle$ is expected to increase linearly with $k$.
In Figure~\ref{FIG4}(D-E) we compare the experimental data with two reference sequences: (i) a Poisson process of the same mean rate as the experimental data (black circles) representing independent and randomly timed events, and (ii) a perfectly regular, equally spaced sequence (stars) representing maximal regularity (details in \SIapp).

Across all orders, the experimental curves lie between these bounds: below the Poisson baseline (indicating stronger correlations than random timing) but above the perfectly regular sequence (indicating retained variability). 
Furthermore, a clear dependence on the window length can be observed: 
For short windows, the moments are close to the equally spaced reference, indicating that, at this timescale, Path B selections are highly regular due to the stranger-following effect (i.e., micro-scale correlation). As the window length increases, the moments gradually move toward the Poisson baseline, as multiple avalanches and periods of inactivity average out short-timescale correlations and become less sensitive to bursty fluctuations (i.e., macro-scale averaging). 
This trend highlights a scale-dependent structure: Path selection is temporally correlated at short timescales but appears more random at longer timescales, a representation of the stranger-following effect.

Overall, these findings reveal that localized social influence triggers collective cascades in routing decisions. Individual choices can rapidly propagate through successive strangers, generating temporally correlated avalanches of path selection. This avalanche behavior underscores the importance of incorporating social contagion into models of pedestrian routing and opens the door to broader discussions on collective dynamics in human mobility.

\section*{Conclusions and Outlook}
In this work, we investigated pedestrian routing behavior in a realistic transit environment, combining high-resolution trajectory measurements, statistical analysis, and theoretical modeling. Our results reveal that path selection is shaped not only by spatial constraints (e.g., exit door locations) and levels of congestion but also by short-range interactions between strangers. Notably, we identified a pronounced and robust stranger-following effect, where individuals are likely to choose the same path as the pedestrian immediately ahead, even in the absence of pre-existing social groups and even when this is not the individual optimal choice. 
As a consequence, the resulting flow does not maximize either the individual travel time or the collective throughput: in a theoretical stationary regime with a very large number of pedestrians, the facility would achieve maximal flow at roughly a 50-50 route split, but the finite and transient nature of the pedestrian flows studied here means that this idealized optimum is not realized in practice, and the observed unbalanced usage that benefits neither side instead arises from the short-range imitation. This effect was systematically quantified through conditional path-choice probabilities and factorial-moment analyses, highlighting both microscale correlations and temporally clustered ``avalanches'' of consecutive choices.
Importantly, the associated dynamics reveal that the stranger-following effect can be overwhelming compared with other contributing factors, often dominating the decision-making process at the bifurcation points.
To understand the underlying mechanisms, we developed a cost-based routing model as a supplementary phenomenological representation, incorporating stochastic walking speeds, herding behavior, and the stranger-following effect. We demonstrated that the stranger-following effect constitutes the dominant driver of route choice at bifurcations, with others serving as secondary modifiers that refine but are insufficient to explain the observed patterns on their own. The strong agreement between model predictions and experimental observations underscores the importance of integrating multi-scale social influences into predictive frameworks for pedestrian dynamics. 
An important methodological insight from the current work is that group ratio can be inferred from path-choice statistics, providing a complementary approach to trajectory-based group detection. 
This is particularly useful in contexts where explicit identification of social groups is challenging and offers an elegant complementary method and cross-check for traditional methods.

The current work also resolves limitations encountered in earlier studies of routing choices \cite{gabbana2021modeling,gabbana2022fluctuations}, where limited statistics and highly imbalanced path usage made it impossible to probe correlations between consecutive choices. In contrast, our long-term recordings yield high statistical resolution and enable us to detect subtle differences in choice probabilities down to the level of a few percent. Crucially, this allows us to uncover a phenomenon entirely absent in earlier studies: the ``follow-stranger effect''. 

Looking ahead, our findings have several implications. First, they highlight the need to account for short-range social contagion in models of pedestrian dynamics, particularly in high-density and safety-critical environments, e.g., train stations or event venues. Second, the avalanche-like temporal clustering of decisions suggests that local interventions (e.g., guidance signage or real-time crowd information) could propagate through the stranger-following effect to influence the collective flow patterns efficiently. More broadly, these results emphasize that spontaneous imitation can drive the system into states that are suboptimal both for individuals and for the facility as a whole, a key consideration for real-time crowd management and infrastructure design. Finally, the combination of empirical, statistical, and modeling approaches developed here provides a versatile framework to study social influence in other types of collective human behavior, ranging from pedestrian flows in urban areas to decision-making evacuation scenarios.

Collective decision-making is often understood to arise through two main mechanisms: a top-down process, where a few designated leaders or individuals drive the behavior of the followers \cite{dyer2009leadership,pelechano2006modeling,boos2014leadership,zhang2021experimental,dyer2009leadership, qingge2007simulating}; and a bottom-up process, where large-scale patterns emerge from local interactions without any designated leadership \cite{couzin2005effective,ding2020experimental}.
Our findings suggest that the collective patterning in pedestrian dynamics offers an intermediate perspective. Specifically, while local social imitation is inherently bottom-up, it effectively amplifies and cascades specific and potentially subtle ``preliminary decisions'' made by a few individuals (e.g., those moving slightly ahead or showing stronger preferences for a specific path), ultimately resulting in macroscale directional guidance. This mechanism reveals how ``leadership'' can be transiently and dynamically ``constructed'' in the local environment through behavioral imitation and decision cascade, even in the absence of formally pre-assigned leaders or centralized control. 

Beyond pedestrian dynamics, the discovery that microscopic social imitation generates macroscopic collective patterns resonates with key questions in multiple disciplines. A similar amplification of local interactions is observed in biological systems \cite{manfrin2006parallel,krause2000fish,parrish1999complexity,sumpter2010collective,couzin2003self,kok2016crowd}, such as ants reinforcing pheromone trails during foraging, fish schools or bird flocks propagating directional preferences, and bacterial or cellular populations responding collectively to chemotactic gradients. In these cases, locally transmitted information, whether chemical, visual, or mechanical, is capable of cascading through the group and producing large-scale organizational patterns. Analogous phenomena appear in active-matter systems \cite{ramaswamy2010mechanics,corbetta2023physics,bacik2025order}, where simple interaction rules lead to emergent patterns (e.g., spontaneous alignment, clustering, or flow redirection). By providing a quantitative framework for how local imitation shapes routing decisions in human crowds, our results create a bridge between empirical pedestrian dynamics and these broader classes of collective behavior, suggesting that conceptually similar mechanisms may emerge across living, technological, and physical systems, even if driven by different underlying factors.

\section*{Acknowledgments}
We would like to thank O. van Erk for his contribution during the early stages of the data analysis. We also acknowledge  C. Pouw and A. Corbetta for useful discussions. This work is partially supported by the HTSM research program ``HTCrowd: a high-tech platform for human crowd flows monitoring, modeling and nudging'' with project number 17962 financed by the Dutch Research Council (NWO).

\section*{Author contributions}
Z.W., A.G, and F.T. designed and performed research; Z.W. analyzed data; and Z.W., A.G, and F.T. wrote the paper.

\clearpage
\section*{\Large{Methods}}

\subsection*{Experimental campaign}

We analyze a dataset comprising pedestrian trajectories collected over a three years period, from March 2021 to March 2024, totaling about 30,000,000 trajectories (including all the disembarking passengers and the people already present on the platform). The data was gathered at tracks 3 and 4 of Eindhoven Centraal Railway Station (The Netherlands) using a commercial, anonymous tracking system based on 3D stereoscopic imaging. This system employs 22 overhead pedestrian tracking sensors (Xovis), covering an area of approximately $1400~\text{m}^2$. Each sensor captures images at a rate of 10 frames per second, processes the stereo images in real-time, and records only the pedestrian locations as $x$, $y$ coordinate pairs. The system uses overhead depth sensing and produces fully anonymized trajectories, without recording RGB images or any identifiable data, in full respect of individual privacy.

An example dataset spanning two months is available in Ref.~\cite{pouw_2024_ehvdata}. 
In our study we specifically focused on pedestrian flows moving toward the platform exit and disembarking 
from trains at designated door locations from track 4 (see Fig.~\ref{FIG1}). 

We apply filters to include only trajectories that satisfy the following criteria: (i) passengers who disembark from track 4 at specific door locations (i.e., $L_1$, $L_2$, and $L_3$), and (ii) trips for which the corresponding train information (arrival and departure times) is complete. 
After filtering, the dataset includes approximately $100,000$ individual trajectories used for our analysis.

Because the stopping position of the train fluctuates slightly between arrivals, the longitudinal positions of the doors vary within several meters. To accommodate this variability, we define the three door ranges as: (i) $L_1: x \in 47.5 \pm 3.75~\text{m}$, (ii) $L_2: x \in 56.25 \pm 5.00~\text{m}$, and (iii) $L_1: x \in 66.25 \pm 5.00~\text{m}$. In the analyzed dataset, approximately 48\% of the trains show two consecutive doors within 7.5~m, 53\% within 10~m, and 79.4\% within 15~m. These statistics indicate that different doors of the same train can fall into different ranges, leading to natural fluctuations in the inflow to Path A/B zones from each door range.

Our analysis does not involve personal data and was approved with Ethical Review ERB2020AP1.

\subsection*{Group detection criterion: average
coexistence distance}
For pedestrians $i$ and $j$, the average
coexistence distance is calculated as 
\begin{equation}
\left\langle d_{i,j} \right\rangle_t=\frac{1}{\Delta t_{i,j}}
\int_{t_{i,j}^{\text{start}}}^{t_{i,j}^{\text{end}}}\!\!\bigl\lVert \vec{x}_i(t)-\vec{x}_j(t) \bigr\rVert\,dt,
\end{equation}
with $\Delta t_{i,j}=t_{i,j}^{\text{end}}-t_{i,j}^{\text{start}}$ and $t_{i,j}^{\text{start}}=\max(t_i^{\text{start}},t_j^{\text{start}}),\;
t_{i,j}^{\text{end}}=\min(t_i^{\text{end}},t_j^{\text{end}})$.

\subsection*{Routing model}
The routing model was built on the cost-based framework from a previous work~\cite{gabbana2021modeling}, in which each pedestrian selects the path with the minimal perceived cost. We extend this model by incorporating two key mechanisms: social influence and stochasticity.

At the bifurcation, each passenger selects a path by minimizing a perceived cost
\begin{equation}
T_i = f_i \cdot r_i^{(\chi)} \cdot S_i \cdot t_i 
= f_i \cdot r_i^{(\chi)} \cdot S_i \cdot \frac{l_i}{V_i},
\end{equation}
where $t_i = l_i/V_i$ is the deterministic travel time. This formulation allows behavioral modifiers to modulate pure time minimization. 
The walking distance $l_i$ of pedestrian $i$ depends on both the chosen path (Path A or Path B) and the door location. Here, we define $l_i$ as the mean streamline length connecting the corresponding door region to the selected path, computed from the spatially binned mean velocity field (see SI for details). 

The first modifier is speed stochasticity, modeled as
\begin{equation}
\label{eq:stoch}
S_i = \frac{V_i}{V_i^{(\text{Stoch})}}, \quad V_i^{(\text{Stoch})} \sim \mathcal{N}(V_i, \sigma_V^2),
\end{equation}
where $V_i^{(\text{Stoch})}$ represents a stochastic realization of walking speed, with $V_i$ estimated from the linear fit in Fig.~\ref{FIG3}A and variance $\sigma_V^2$ estimated empirically from the full experimental dataset ($\sigma_V = 0.116\,\mathrm{m/s}$). This term captures the natural heterogeneity in walking speed across individuals, introducing variability in predicted costs even under identical crowding conditions.  

The second modifier accounts for herding behavior. We introduce a multiplicative penalty function \cite{gabbana2021modeling}
\begin{equation}
\label{eq:penalty}
f_i = (1+a) - a \cdot \tanh\!\left[s \cdot (n_i - n_0)\right],
\end{equation}
where $a$, $s$, and $n_0$ are tunable parameters, and $n_i = \frac{N_i}{N_A+N_B}$ is the fraction of pedestrians present along the path that is chosen by the $i^{\text{th}}$ pedestrian at the time of decision \cite{gabbana-cd-2021}. This function penalizes minority choices and rewards majority-following decisions.
The hyperbolic tangent form ensures smooth saturation of the penalty as group imbalances grow, reflecting diminishing marginal influence when one path becomes strongly dominant.  

Finally, we incorporate a reward to capture short-range imitation of the immediately preceding pedestrian, i.e., stranger-following effect. This is defined as
\begin{equation}
\label{eq:reward}
r_i^{(\chi)} = 1 - \gamma \cdot \mathbb{I}(\text{Prev}_i = \chi),
\end{equation}
where $\gamma \in [0,1]$ tunes the reward magnitude, and $\mathbb{I}(\cdot)$ is an indicator function equal to 1 if the current pedestrian chooses the same path $\chi$ as the previous pedestrian, and 0 otherwise. A reduced cost thus arises when successive pedestrians align in their choices, effectively biasing decision-making toward local consistency and transient stranger-following dynamics.

\clearpage
\onecolumngrid 

\renewcommand{\figurename}{\textbf{Supplementary Figure}}
\renewcommand{\theequation}{S.\arabic{equation}}
\setcounter{figure}{0} 
\setcounter{equation}{0}

\section*{\Large{Supplementary Information}}

\section{Comparison between stranger-to-stranger imitation and traditional follow-the-leader behavior}
To clarify the mechanism underlying our observations, we summarize the conceptual differences between the classical follow-the-leader paradigm and the stranger-to-stranger imitation reported in our study (Table~\ref{SI_tab_follow}).

Unlike classical leader-following, the stranger-to-stranger imitation we observe:
\begin{enumerate}
    \item Emerges among anonymous strangers without social hierarchy or communication;
    \item Is purely local and transient, based on the immediate predecessor only;
    \item Arises because an individual may implicitly infer that the preceding pedestrian may have a locally better reason to choose a route under uncertainty.
\end{enumerate}
Despite its minimal nature, this mechanism is sufficient to generate bursty cascades of route choices.
\section{Definition of the order of the exiting passengers}
\begin{figure}
\centering
\includegraphics[width=0.9\textwidth]{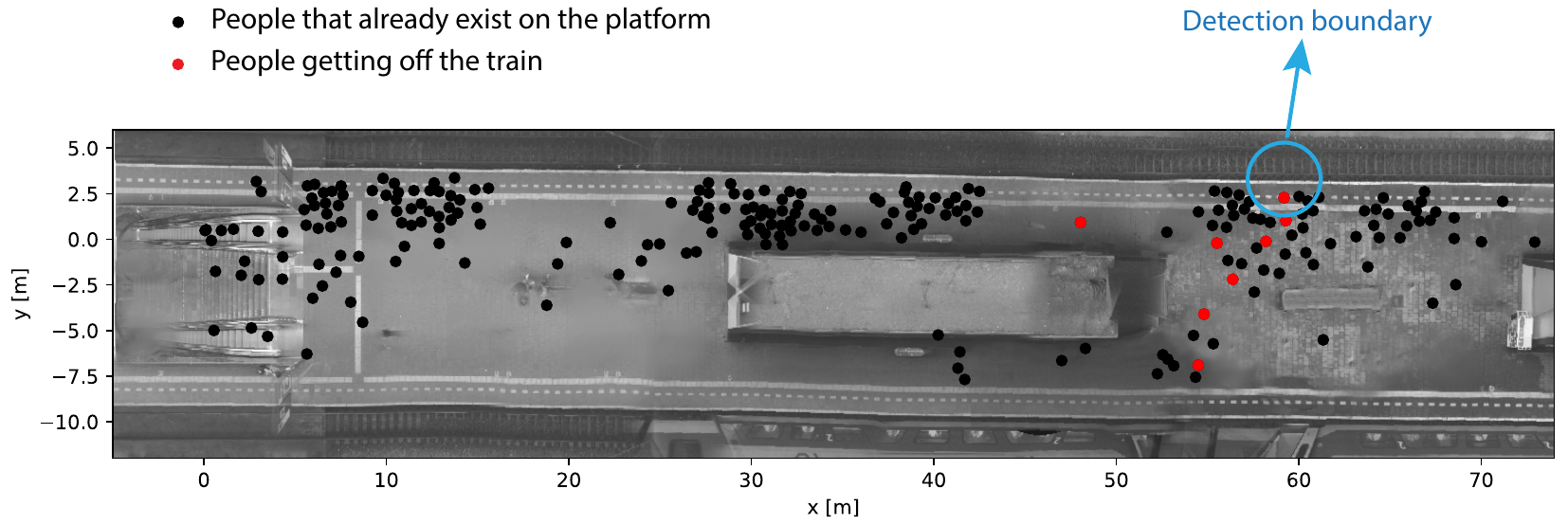}
\caption{\textbf{Definition of the order of the exiting passenger.} The black dots denote the people who have already standing on the platform. The red dots denote the passengers who just get off the train. The blue circle represents the detection boundary based on which the order of exiting passenger is defined.}
\label{fig:order_definition}
\end{figure}

In this study, the order of the exiting passengers is defined strictly by the temporal order in which passengers exit the train. 
The typical width of a train door is $1-1.3$ m (information obtained from \url{trainsforumnl.jouwweb.nl} and \url{https://www.treinenweb.nl/materieel/VIRM/}), which effectively allows only one passenger to pass through at a time. The initial appearance of a passenger in the camera view can exhibit small unstable oscillations, the exact moment of detection may vary. To avoid this temporal detection uncertainty when determining the exit order of passengers, 
around each door for each arriving train, a two-meter circular detection boundary is defined (see Fig.~\ref{fig:order_definition}). Any individual who first appears within this boundary during the interval $(t_0, t_0 + \Delta t)$, where $\Delta t = \min\{t_1 - t_0,~300~\text{s}\}$, is identified as a passenger exiting from that door, with $t_0$ the train arrival time and $t_1$ the train departure time, and they will form the dataset for analyzing the route choices. The time at which a pedestrian first crosses the boundary is recorded as their exit time. The sequence of exiting passengers is therefore established solely based on these exit times, and the preceding pedestrian for passenger $i$ is defined as the individual who exits immediately before them. This procedure eliminates ambiguity arising from two-dimensional motion, local proximity changes, or field-of-view variations.

\section{Generalized congestion measure}
Congestion on Path A (Path B) is quantified by the instantaneous number of pedestrians, $N_A$ ($N_B$), located in the Path A (Path B) region at the moment a new passenger exits the train.
This measure reflects the combined effects of several factors, including
(i) the number of passengers descending from different door ranges,
(ii) the number of pedestrians already present in the area, and
(iii) stochastic variations in walking speed and delays.
Therefore, variations in the inflow from any particular door range are naturally incorporated into this aggregated congestion indicator.
Although door-level inflows contribute to fluctuations in Path A occupancy, the observed sequential dependence in route choice (i.e., $P(X_i=\chi \mid X_{i-1}=\chi, \neg G_i^{(2)}) \approx P(X_i=\chi \mid X_{i-1}=\chi) > P(X_i=\chi)$ reported in Fig.~2(A-C) in the main text) remains robust across all levels of congestion, indicating that the stranger-following effect reflects a general behavioral tendency rather than a condition limited to particular inflow scenarios.

\section{Group detection methodology}
Two pedestrians are considered part of the same group if they coexist for longer than five seconds and satisfy all three conditions:

(i) The time-averaged inter-personal distance during coexistence, $\left\langle d_{i,j} \right\rangle_t$ (details in Methods), is smaller than a chosen threshold. 
For pedestrians $i$ and $j$
\begin{equation}
\left\langle d_{i,j} \right\rangle_t=\frac{1}{\Delta t_{i,j}}
\int_{t_{i,j}^{\text{start}}}^{t_{i,j}^{\text{end}}}\!\!\bigl\rVert \vec{x}_i(t)-\vec{x}_j(t) \bigr\rVert\,dt,
\end{equation}
where $t_{i,j}^{\text{start}}=\max(t_i^{\text{start}},t_j^{\text{start}}),\;
t_{i,j}^{\text{end}}=\min(t_i^{\text{end}},t_j^{\text{end}}),\;
\Delta t_{i,j}=t_{i,j}^{\text{end}}-t_{i,j}^{\text{start}}$.

(ii) The covariance of their speeds, $\mathrm{Cov}(v_i,v_j)$, exceeds a threshold;

(iii) The velocities of the two pedestrians are sufficiently aligned, i.e., the cosine value of the angle $\theta$ between $\vec{v}_i$ and $\vec{v}_j$ exceeds a threshold. 

Based on these criteria, the group with a leader (who exits the train earlier) and several members (one or more) is identified.

\begin{figure}
\centering
\includegraphics[width=0.8\textwidth]{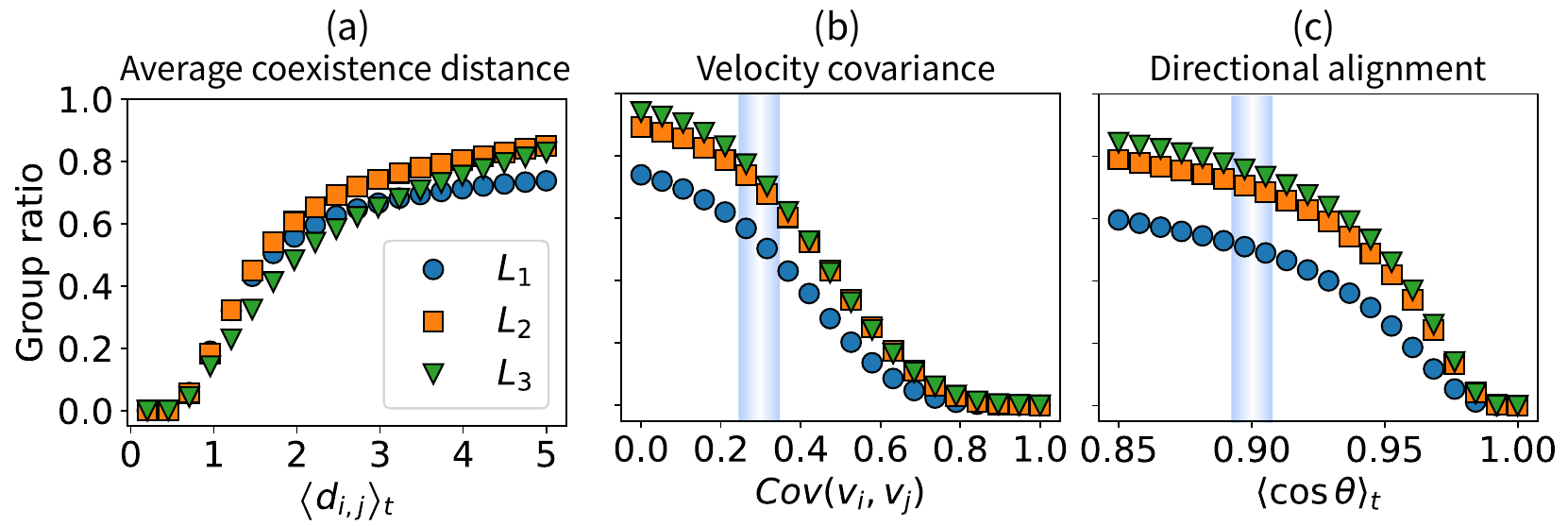}
\caption{\textbf{Group detection criteria.} The detected group ratio as a function of the chosen threshold based on the criteria of (a) average coexistence distance; (b) velocity covariance; and (c) directional alignment for door position of $L_1$ (blue), $L_2$ (orange), and $L_3$ (green), respectively.}
\label{fig:group_detection}
\end{figure}

The detected group ratio, $P(G)$, was examined as it varies with the selected threshold, using three distinct criteria to identify groups. The first criterion, average coexistence distance, measures how close individuals are to one another over time. The second, velocity covariance, captures the degree to which individuals move with similar speed patterns. The third, directional alignment, quantifies how consistently individuals align their movement directions with each other. Together, these criteria provide a comprehensive assessment of how the threshold affects group detection. As shown in Fig. \ref{fig:group_detection}, the detected group ratio is presented as a function of the chosen threshold based on the criteria of (a) average coexistence distance; (b) velocity covariance; and (c)directional alignment for door positions of $L_1$ (blue), $L_2$ (orange), and $L_3$ (green), respectively.

\begin{figure}
\centering
\includegraphics[width=0.5\textwidth]{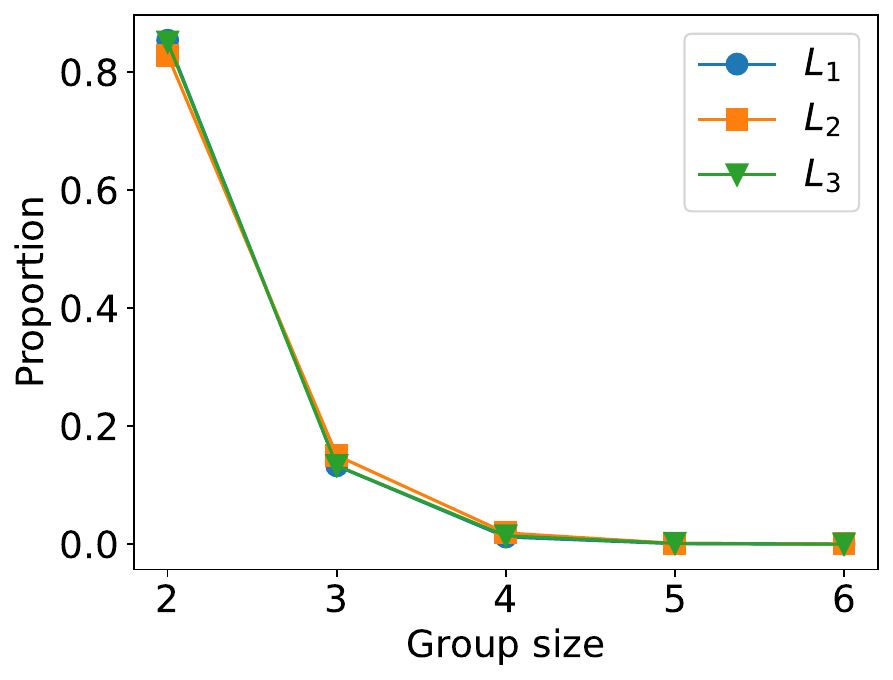}
\caption{\textbf{Detected group size distribution.} The distribution of the detected group size for door position of $L_1$ (blue), $L_2$ (orange), and $L_3$ (green), respectively. For example, group size is three means that in such a group there are two group members following the group leader.}
\label{fig:group_size}
\end{figure}

The distributions of the detected group sizes for the door locations of $L_1$ (blue), $L_2$ (orange), and $L_3$ (green), respectively, are shown in Fig.~\ref{fig:group_size}. The tendency of the group size distribution observed in the current study agrees with that reported in \cite{moussaid2010walking}.


\section{Calculation of walking distance $l_i$}

In the theoretical model, we need to explicitly calculate the walking distance $l_i$ of pedestrian $i$, whose value depends on both the door location ($L_1$ ($x \in 47.50 \pm 3.75$\,m), $L_2$ ($x \in 56.25 \pm 5.00$\,m), $L_3$ ($x \in 66.25 \pm 5.00$\,m)) and the choice between Path A and Path B. To ensure consistency, we define $l_i$ as the mean length of the streamlines obtained by integrating the spatially binned
mean velocity field for pedestrians exiting from door $L_i$. 
Specifically, for each door $L_i$, we vary the starting points within the door range to generate streamlines leading to Path A and Path B. The average length of all streamlines reaching Path~A gives $l_i$ for Path A, and the average length of those reaching Path~B gives $l_i$ for Path B.

We illustrate the calculation by considering a pedestrian exiting from door $L_1$ and selecting Path A.  
The procedure is as follows:

\begin{enumerate}
\item \textbf{Average streamlines}
The observed walking area is divided into a regular spatial grid, and the mean pedestrian velocity is computed in each cell.  
Individual trajectories are extracted from the depth‐camera recordings and filtered to retain only those originating from $L_1$.  
The trajectories are smoothed using a Savitzky–Golay filter \cite{savitzky1964smoothing} to reduce noise and suppress discontinuities that could otherwise affect velocity estimates.  
Instantaneous velocities are then obtained by finite differencing of consecutive smoothed positions.  
For each spatial bin, the velocity vectors of all relevant trajectory segments are averaged to yield the local mean velocity, stored as two matrices 
$\bar{v}_x$ and $\bar{v}_y$ for the $x$- and $y$-components, respectively.  
A two–dimensional mean velocity field is visualized by integrating streamlines through these matrices, producing continuous curves indicating the dominant flow, as shown in Fig.~1(B–D) of the main text and in Figs.~\ref{SI_fig_L1_cal_li_traj}–\ref{SI_fig_L3_cal_li_traj}.

\item \textbf{Define starting points and velocity field interpolation}
To generate a streamline from an arbitrary starting location within the door region, the velocity vector 
$\vec{v}(x(t),y(t))$ at any position is obtained by interpolating $\bar{v}_x$ and $\bar{v}_y$.  
The streamline is then integrated using \cite{kundu2024fluid}
$$
\frac{dx}{dt} = v_x(x(t),y(t)), \qquad
\frac{dy}{dt} = v_y(x(t),y(t)),
$$
with the initial point $(x_0,y_0)$ chosen within the spatial range of the given door.

\item \textbf{Average distance}
By varying the starting point across the $L_1$ door range ($x \in 47.50 \pm 3.75$\,m), multiple streamlines are generated (examples shown as thick solid lines in Figs.~\ref{SI_fig_L1_cal_li_traj}–\ref{SI_fig_L3_cal_li_traj}).  
The length of each streamline is measured, and the mean of these lengths defines $s_i$ for pedestrians exiting from $L_1$ along Path A.

\item \textbf{Apply consistently across doors and paths.}  
The same procedure is repeated for each door–path combination to obtain values of $s_i$ across the dataset.
\end{enumerate}

To resolve the streamlines more accurately near the doors, we focus on the region $x \in [40,70]$\,m,
which includes all three door locations.  
The Path A region spans $x \in [29,49.5]$\,m, $y \in [-0.4,3.7]$\,m,  
and the Path B region spans $x \in [29,49.5]$\,m, $y \in [-9.2,-5.1]$\,m.  
Within $x \in [29,40]$\,m, the travel path is approximately straight with a length of $\Delta s = 11$\,m.  
Hence, the total walking distance is given by
$$
l_i = s_i + \Delta s.
$$

\begin{figure}
\centering
\includegraphics[width=\textwidth]{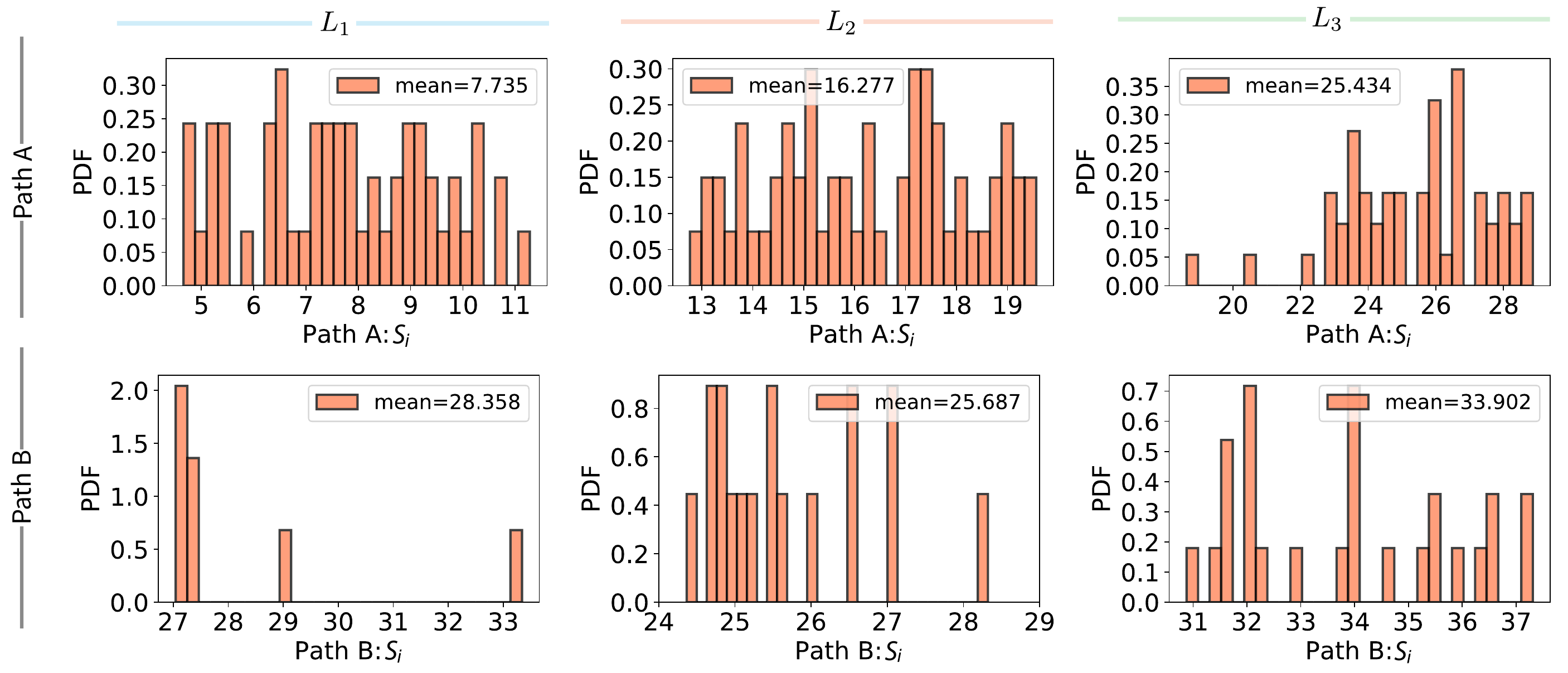}
\caption{PDF of the travel distance of the average streamline along Path A (top row) and Path B (bottom row) for pedestrians exiting from door $L_1$ (left column), $L_2$ (middle column), and $L_3$ (right column), respectively.}
\label{SI_fig_cal_li_pdf}
\end{figure}

\begin{figure}
\centering
\includegraphics[width=\textwidth]{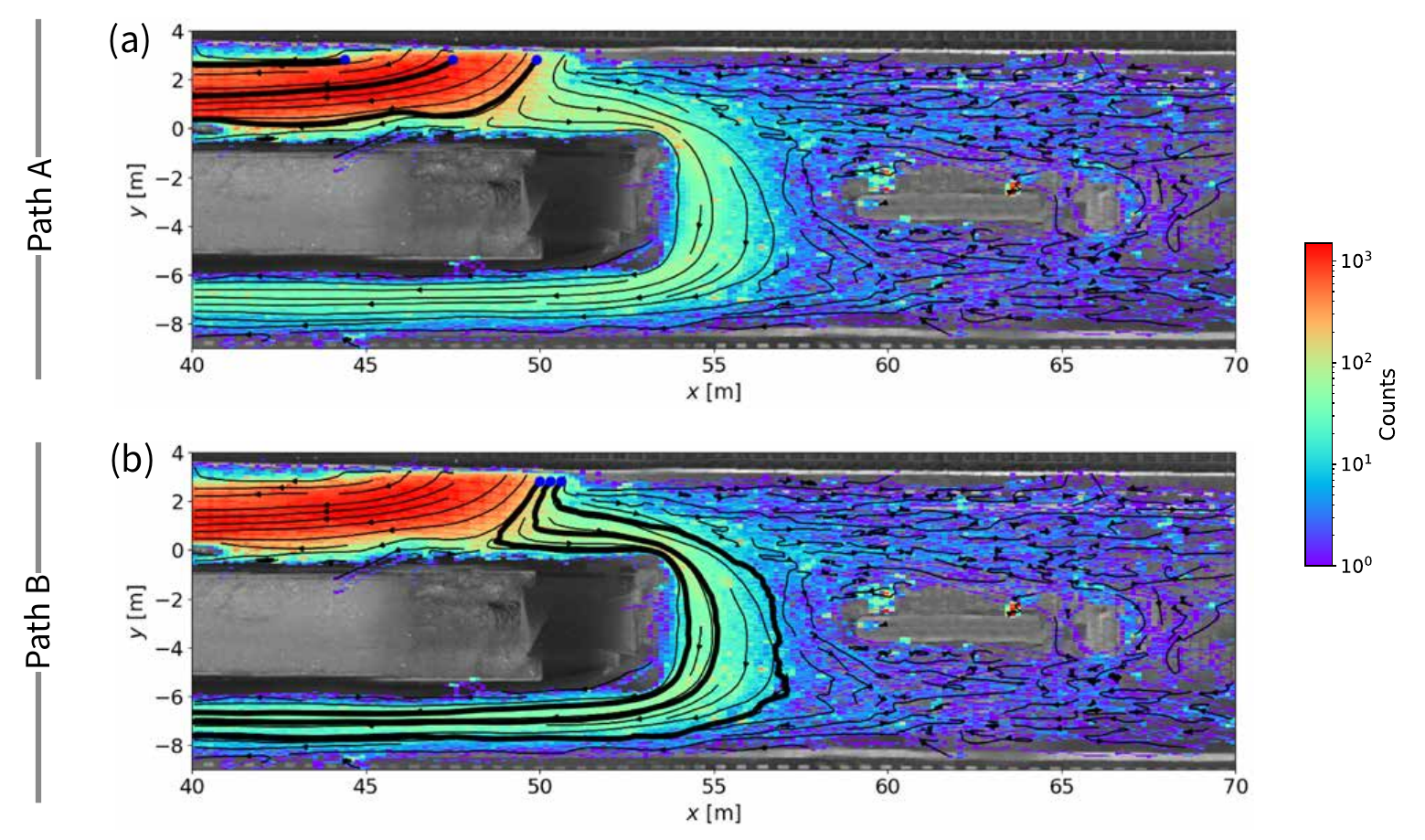}
\caption{Heat maps of pedestrian positions after exiting through doors $L_1$. Colors represent the density of observed positions on a logarithmic scale based on a 2D count histogram. Overlaid streamlines show the mean velocity vector field (spatially binned), highlighting the most probable pedestrian trajectories.
Thick solid lines: representative average streamlines along Path A (a) and Path B (b). The dark blue circles mark the starting location, whose coordinates are $[44.4, 2.8]$ (leftmost), $[47.5, 2.8]$ (middle), and $[49.9, 2.8]$ (rightmost), respectively for (a); and $[50.0, 2.8]$ (leftmost), $[50.3, 2.8]$ (middle), and $[50.6, 2.8]$ (rightmost), respectively for (b).}
\label{SI_fig_L1_cal_li_traj}
\end{figure}

\begin{figure}
\centering
\includegraphics[width=\textwidth]{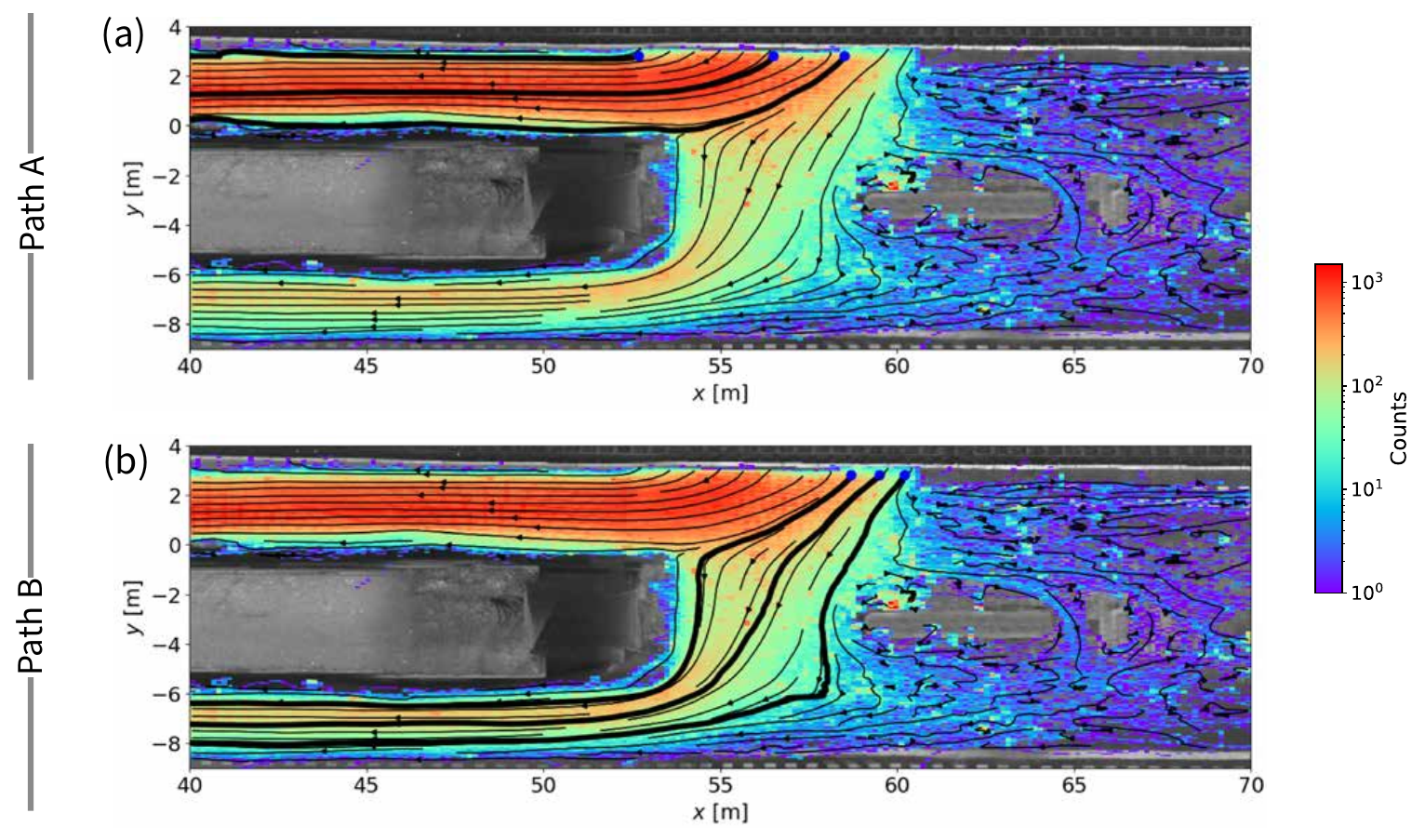}
\caption{Heat maps of pedestrian positions after exiting through doors $L_2$. Colors represent the density of observed positions on a logarithmic scale based on a 2D count histogram. Overlaid streamlines show the mean velocity vector field (spatially binned), highlighting the most probable pedestrian trajectories.
Thick solid lines: representative average streamlines along Path A (top) and Path B (bottom). The dark blue circles mark the starting location, whose coordinates are $[52.7, 2.8]$ (leftmost), $[56.5, 2.8]$ (middle), and $[58.5, 2.8]$ (rightmost), respectively for (a); and $[58.7, 2.8]$ (leftmost), $[59.5, 2.8]$ (middle), and $[60.2, 2.8]$ (rightmost), respectively for (b).}
\label{SI_fig_L2_cal_li_traj}
\end{figure}

\begin{figure}
\centering
\includegraphics[width=\textwidth]{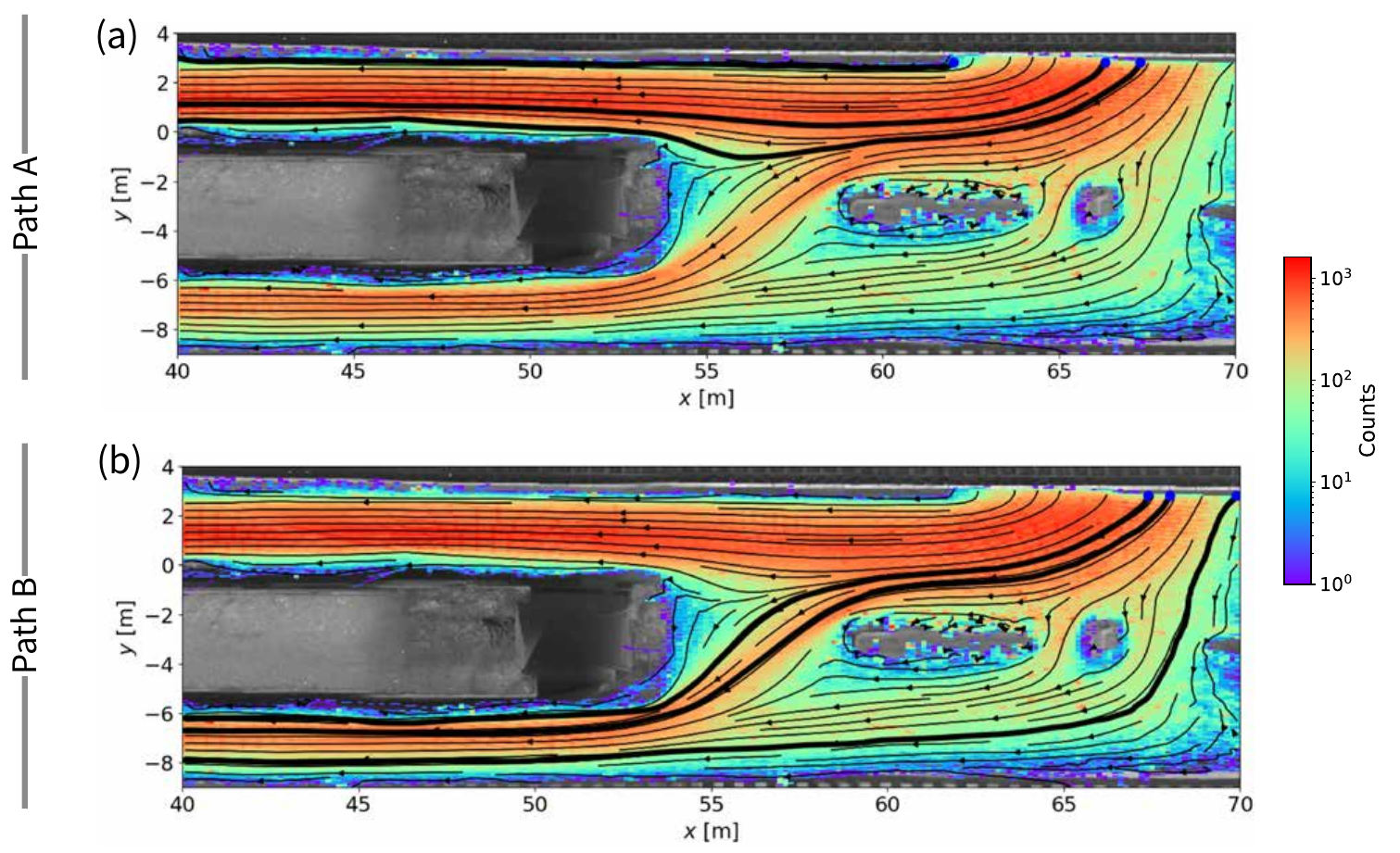}
\caption{Heat maps of pedestrian positions after exiting through doors $L_3$. Colors represent the density of observed positions on a logarithmic scale based on a 2D count histogram. Overlaid streamlines show the mean velocity vector field (spatially binned), highlighting the most probable pedestrian trajectories.
Thick solid lines: representative average streamlines along Path A (top) and Path B (bottom). The dark blue circles mark the starting location, whose coordinates are $[62.0, 2.8]$ (leftmost), $[66.3, 2.8]$ (middle), and $[67.3, 2.8]$ (rightmost), respectively for (a); and $[67.4, 2.8]$ (leftmost), $[68.0, 2.8]$ (middle), and $[69.9, 2.8]$ (rightmost), respectively for (b).}
\label{SI_fig_L3_cal_li_traj}
\end{figure}

Figure~\ref{SI_fig_cal_li_pdf} presents the probability density function (PDF) of $s_i$—the average streamline length within $x \in [40,70]$\,m—for Path~A (top row) and Path~B (bottom row) for pedestrians exiting from doors $L_1$, $L_2$, and $L_3$.  
The resulting mean travel distances $l_i$ are summarized in Table~\ref{tab:li}.

\section{Inferring group ratio from path-choice statistics: detailed derivation}
We have mentioned in the main text that the group ratio, $P(G)$, can be directly inferred from path-choice statistical characteristics. 

To formalize this, the individual ($I_i$) and the group leader ($G_i^{(1)}$) are classified into the same type, i.e., $\neg G_i^{(2)}$. Under this classification, the conditional probability that pedestrian $i$ repeats the choice made by the immediately preceding pedestrian can be decomposed as \cite{durrett2019probability},
\begin{equation}
\begin{aligned}
&P(X_i = B \mid X_{i-1}=B) = \\
&P(X_i = B \mid X_{i-1}=B, G_i^{(2)}) \cdot P(G_i^{(2)} \mid X_{i-1}=B) + \\
&P(X_i = B \mid X_{i-1}=B, \neg G_i^{(2)}) \cdot P(\neg G_i^{(2)} \mid X_{i-1}=B)
\end{aligned}
\label{eq:total_prob_law}
\end{equation}

To separate the effects of group and independent behavior, we note that for any pedestrian $i$ following a previous choice $X_{i-1} = B$, the probability of belonging to a group or not sums to unity:
\begin{align}
    P(G_i^{(2)} \mid X_{i-1}=B) + P(\neg G_i^{(2)} \mid X_{i-1}=B) = 1.
\label{eq:prob_unity}
\end{align}

Combining Eq.~\ref{eq:total_prob_law} and Eq.~\ref{eq:prob_unity}, we can solve for the conditional probability that pedestrian $i$ belongs to a group:
\begin{equation}
\begin{aligned}
    &P(G_i^{(2)} \mid X_{i-1}=B) =\\ 
    &\frac{
    P(X_i=B \mid X_{i-1}=B) - P(X_i=B \mid X_{i-1}=B, \neg G_i^{(2)})
    }{
    P(X_i=B \mid X_{i-1}=B, G_i^{(2)}) - P(X_i=B \mid X_{i-1}=B, \neg G_i^{(2)})
    }
\end{aligned}
\label{eq:cal_group_ratio_B}
\end{equation}

Based on the assumption that two pedestrians from the same group always choose the same path, we have 
\begin{equation}
P(X_i=B \mid X_{i-1}=B, G_i^{(2)}) = 1. 
\end{equation}

This is because if the pedestrian $i$ is a group member, then the pedestrian $(i-1)$ must be the group leader. Since conditioning on $X_{i-1}=B$ implies that the group leader has chosen $B$, and the group member necessarily does the same, i.e., $X_i=B$.

Eq.~\ref{eq:cal_group_ratio_B} then simplifies to
\begin{equation}
\begin{aligned}
    &P(G_i^{(2)} \mid X_{i-1}=B) =\\ 
    &\frac{
    P(X_i=B \mid X_{i-1}=B) - P(X_i=B \mid X_{i-1}=B, \neg G_i^{(2)})
    }{
    1 - P(X_i=B \mid X_{i-1}=B, \neg G_i^{(2)})
    }
\end{aligned}
\label{eq:cal_group_ratio_B2}
\end{equation}
where all terms on the right-hand side have been directly measured from the experimental data (blue in Fig.~2(A-C) in the main text).

A completely analogous expression holds for path A:
\begin{equation}
\begin{aligned}
    &P(G_i^{(2)} \mid X_{i-1}=A) =\\ 
    &\frac{
    P(X_i=A \mid X_{i-1}=A) - P(X_i=A \mid X_{i-1}=A, \neg G_i^{(2)})
    }{
    1 - P(X_i=A \mid X_{i-1}=A, \neg G_i^{(2)})
    }
\end{aligned}
\label{eq:cal_group_ratio_A}
\end{equation}
with the corresponding inputs taken from Fig.~a(A-C) (green) of the main text.

Now we can express the overall probability of a pedestrian $i$ being a group member using the law of total probability \cite{durrett2019probability}:
\begin{equation}\label{eq:cal_pG}
\begin{split}
    P(G_i^{(2)}) = &P(G_i^{(2)} \mid X_{i-1}=A) \cdot P(X_{i-1}=A)+ \\
                   &P(G_i^{(2)} \mid X_{i-1}=B) \cdot P(X_{i-1}=B)
\end{split}
\end{equation}

This can be equivalently written as
\begin{equation}
\begin{split}
    P(G_i^{(2)}) = &\, P(G_i^{(2)} \mid X_{i-1}=A) \cdot P(X_i=A) \\
                   &+ P(G_i^{(2)} \mid X_{i-1}=B) \cdot P(X_i=B)
\end{split}
\end{equation}
where $P(X_{i}=A)$ and $P(X_{i}=B)$ are obtained from experimental results, as shown in Fig.~2I in the main text.

Finally, noting that each group consists of two members, the estimated group ratio is
\begin{align}
    P(G) = P(G_i^{(2)}) * 2
\end{align}

\section{Velocity probability density function (PDF) for two different density ranges}
\begin{figure}
\centering
\includegraphics[width=0.6\textwidth]{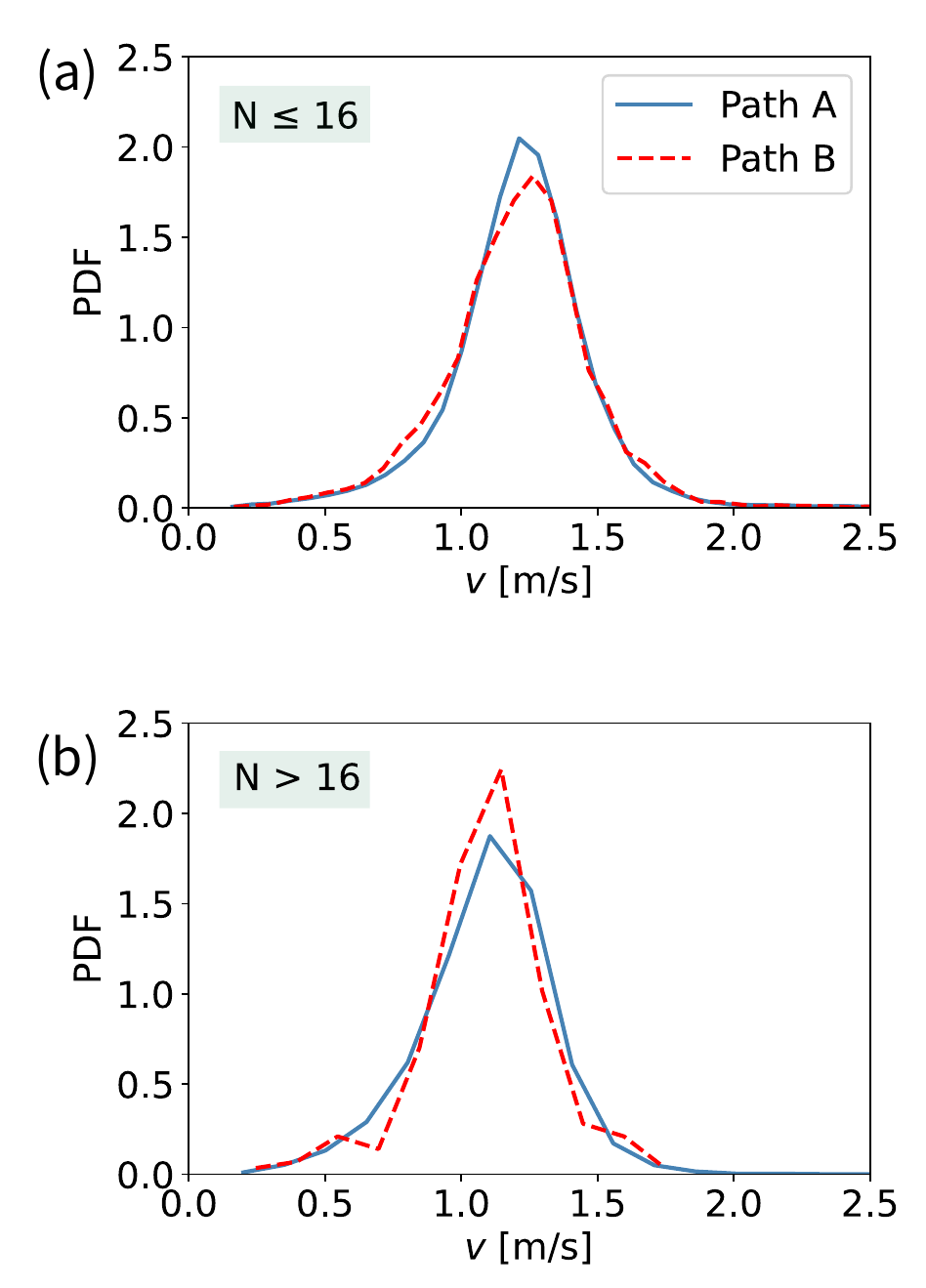}
\caption{Velocity probability density function (PDF) for two different density ranges: (a)$N \le 16$, and (b) $N>16$.}
\label{SI_fig_v_PDF}
\end{figure}

The probability function (PDF) for pedestrian velocity is estimated for two crowd density ranges: (i) $N \leq 16$; and (ii) $N>16$.
As shown in Fig.~\ref{SI_fig_v_PDF}, the blue line is for Path A, and the red-dashed line is for Path B. 

\section{Extra model analysis}
\begin{figure}
\centering
\includegraphics[width=0.99\textwidth]{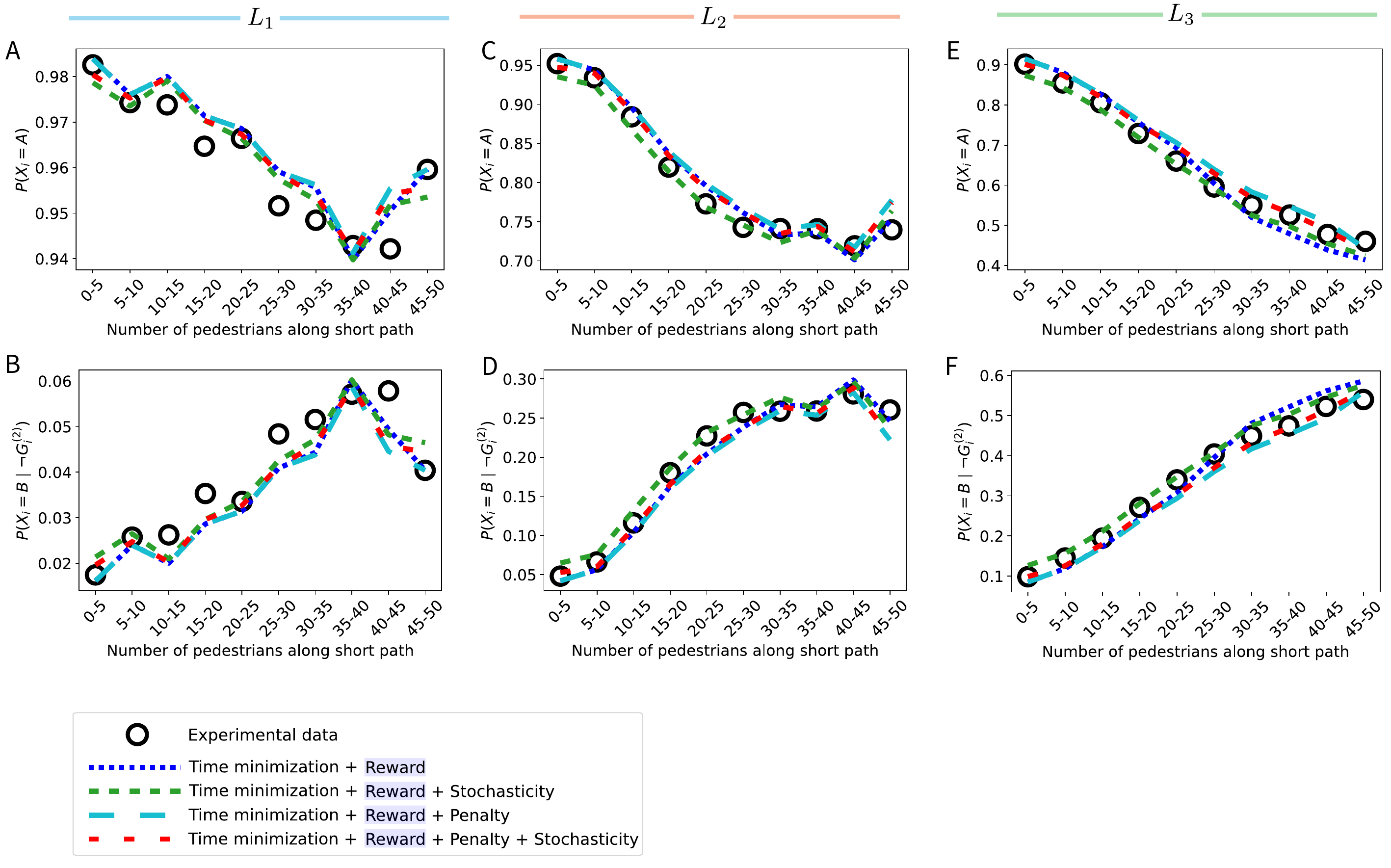}
\caption{Model-data comparison of the conditional probability excluding groups of choosing Path A (panels A, C, E) and Path B (panels B, D, F) as a function of the number of pedestrians on Path A, for exit doors $L_1$, $L_2$, and $L_3$.
Black symbols: experiments. 
Blue dotted: following stranger only ($r_i^{(\chi)} \cdot t_i$).
Green dotted: following stranger $+$ speed variability.
Light-blue dashed: following stranger $+$ herding ($r_i^{(\chi)} \cdot f_i \cdot t_i$).
Red dashed: full model ($f_i \cdot r_i^{(\chi)} \cdot S_i \cdot t_i$).
}
\label{SI_FIG_extra_model}
\end{figure}

In the main paper, we have demonstrated that, when starting from a purely time-minimization baseline, neither stochasticity nor penalty terms alone are sufficient to reproduce the experimentally observed routing behavior. In contrast, introducing the reward term correctly predicts the experimental trend. Here, by presenting extra model prediction results, we show that any additional modifiers only lead to minor adjustments. 
As shown in Fig.~\ref{SI_FIG_extra_model}, models that include extra components (Reward + Stochasticity, Reward + Penalty, Reward + Both) yield similar predictions to that from the Reward-only model. 
Among all model components, the reward term is the decisive factor responsible for reproducing the experimental trend. Stochasticity and penalty functions merely provide small refinements and are included primarily for completeness and for comparison with standard formulations in literature \cite{gabbana2021modeling,gabbana2022fluctuations}. Notably, they do not alter the qualitative trend predicted by the Reward-only model.

\section{Calculation of the effective local decay rate $\lambda(\Delta t)$}
To evaluate the temporal dependence of the effective rate $\lambda(\Delta t)$ associated with the distribution of inter-person time intervals, $\Delta t$, for pedestrians choosing Path B, we adopted a local fitting procedure over the tail of the empirical probability density function (PDF). After excluding short time intervals $\Delta t<0.8$ s, where geometric and social constraints bias the statistics, we fitted the decay of the PDF in a semi-logarithmic representation. Specifically, the logarithm of the histogram counts $\log P(\Delta t)$ was regressed against $\Delta t$ using sliding windows of fixed length.
\\
Let $x$ denote the bin centers and $y= \log P(\Delta t)$ the corresponding logarithmic counts. For each window of width approximately $7.5$ s, containing 20 consecutive bins, we performed a linear regression of the form 
\begin{equation}
    \log P(\Delta t) = a - \lambda_{\text{loc}} \Delta t,
\end{equation}
which yields a local estimate of the decay rate $\lambda_{\text{loc}}$. The window was then shifted by approximately $1.13$ s, and the fitting procedure was repeated. The resulting values were assigned to the midpoint of each window, producing an effective rate $\lambda(\Delta t)$ as a function of $\Delta t$.
\\
This sliding-window fit ensures that (i) each estimate is obtained from a sufficiently large sample (20 histogram bins), (ii) the analysis remains robust to local fluctuations, and (iii) potential deviations from a constant Poissonian rate can be directly quantified.
\\
The resulting $\lambda(\Delta t)$ allows us to distinguish between a homogeneous Poisson process and an interaction-driven propagation of choices. 
\section{Factorial moment analysis}
The factorial moment of order $k$ is defined as
\begin{equation}
    \left< n^{(k)} \right> = \mathbb{E}\!\left[ X^{\underline{k}} \right]
= \mathbb{E}\!\left[ X (X-1)(X-2)\cdots (X-k+1) \right],
\end{equation}
where $X^{\underline{p}} = X (X-1)(X-2)\cdots (X-k+1)$ is the falling factorial. For a Poisson distribution, $ \left< n^{(k)} \right> = \lambda^k, \quad p=1,2,3,\ldots$, so the normalized factorial-moments ratio, $\frac{\log\langle n^{(k)} \rangle}{\log\langle n^{(1)} \rangle}$, is expected to be linear in $k$ with a slope of one. 

One of the two reference sequences is a perfectly regular, equally spaced sequence (stars) representing maximal regularity.
In this work, the equally spaced reference sequence is chosen to contain $N=5$ Path B selection events in each sliding window, so that the fifth-order factorial moment exists. The factorial moments of order $p$ are given by $\left< n^{(k)} \right> = N(N-1) \ldots (N-k+1)$. Of course, as the chosen $N$ increases, the slope of the normalized ratio $\log\langle n^{(k)}\rangle / \log\langle n^{(1)}\rangle$ is expected to increase.

\begin{table}[h!]
\centering
\caption{Conceptual distinction between traditional follow-the-leader behavior and the stranger-to-stranger imitation observed in our experiment.}
\label{SI_tab_follow}
\begin{tabular}{p{3.5cm} p{3.2cm} p{4.2cm} p{3.5cm}}
\hline
\textbf{Mechanism} & \textbf{Who is followed?} & \textbf{Information basis} & \textbf{Typical setting} \\
\hline
\textbf{Traditional follow-the-leader} \cite{reebs2000can,dyer2009leadership,pelechano2006modeling,boos2014leadership,ding2020experimental,zhang2021experimental,dyer2009leadership,qingge2007simulating,xie2022simulation,yu2025study} &
A designated or visible leader &
Expectation of knowledge, competence, or authority &
Emergency guidance, trained personnel, animal groups with hierarchy \\
\hline
\textbf{Stranger-to-stranger imitation (this study)} &
Random anonymous pedestrian immediately ahead &
Assumed local preference under uncertainty, despite no verified information &
Crowds with no leaders, no prior knowledge, and no communication \\
\hline
\end{tabular}
\end{table}

\begin{table}[h!]
  \centering
  \caption{Mean travel distance, $l_i$, along Path A and Path B for pedestrians exiting from doors $L_1$, $L_2$, and $L_3$, respectively, to the leftmost of the corresponding Path region.}
  \begin{tabular}{p{2.5cm} p{3.2cm} p{3.2cm} }
    \hline
    Door location & Path A: $l_i$ [m]& Path B: $l_i$ [m]\\ \hline
    $L_1$         & 18.735        & 39.358   \\ \hline
    $L_2$         & 27.277        & 36.687   \\ \hline
    $L_3$         & 36.434        & 44.902   \\ \hline
  \end{tabular}
  \label{tab:li}
\end{table}

\clearpage
\bibliography{main}

\end{document}